\documentclass[preprint,12pt,authoryear]{elsarticle}

\usepackage{mathabx} 

\usepackage{color}
\usepackage{lineno}

\usepackage[colorlinks]{hyperref}
\hypersetup{
  citecolor=blue, 
  linkcolor=blue,
   urlcolor=blue
}

\newcommand{\orb}{{\tt orbitN}}
\newcommand{\snv}{\mbox{\tt snvec}}
\newcommand{\snvR}{\mbox{\tt snvecR}}
\newcommand{\ascr}{\mbox{\tt astrochron}}
\newcommand{\acl}{\mbox{\tt Acycle}}
\newcommand{\rbnd}{\mbox{\tt rebound}}
\newcommand{\myui}{\mbox{\tt astrocyclo}}

\newcommand{\giturl}{\url{github.com/rezeebe/orbitN}}
\newcommand{\myurl}{\url{www2.hawaii.edu/~zeebe/Astro.html}}
\newcommand{\uiurl}{\url{astrocyclo.soest.hawaii.edu}}

\newcommand{\npurlETa} {\url{www.ncdc.noaa.gov/paleo/study/26970}}
\newcommand{\npurlXXx} {\url{www.ncdc.noaa.gov/paleo/study/36415}}
\newcommand{\npurlPT}  {\url{www.ncdc.noaa.gov/paleo/study/35174}}
\newcommand{\npurlXXtx}{\url{www.ncdc.noaa.gov/paleo/study/39199}}

\newcommand{\snvurl}{\url{github.com/rezeebe/snvec}}
\newcommand{\snvurlR}{\url{github.com/japhir/snvecR}}

\newcommand{\LaXurl}{\url{vo.imcce.fr/insola/earth/online/earth/La2010/index.html}}
\newcommand{\rbndurl}{\url{rebound.readthedocs.io/en/latest}}
\newcommand{\ascrurl}{\url{cran.r-project.org/web/packages/astrochron/index.html}}
\newcommand{\aclurl}{\url{acycle.org}}

\newcommand{\rev}{\textcolor{black}}

\newcommand{\scs}{\scriptsize}

\newcommand{\AS}{\mbox{$\cal AS$}}
\newcommand{\ASs}{\mbox{$\cal AS$s}}
\newcommand{\calF}{\mbox{$\cal F$}}
\newcommand{\calH}{\mbox{$\cal H$}}
\newcommand{\eE}{\mbox{$e_{\Earth}$}}
\newcommand{\DeE}{\mbox{$\Delta e_{\Earth}$}}
\newcommand{\MS}{\mbox{$M_{\Sun}$}}
\newcommand{\mE}{\mbox{$m_{\Earth}$}}
\newcommand{\mL}{\mbox{$m_{\Moon}$}}
\newcommand{\Ed}{\mbox{$E_d$}}
\newcommand{\Td}{\mbox{$T_d$}}

\newcommand{\ZBSTx}{\mbox{\texttt{ZB17x}}}
\newcommand{\ZBETa}{\mbox{\texttt{ZB18a}}}
\newcommand{\ZBXXx}{\mbox{\texttt{ZB20x}}}
\newcommand{\ZBXXa}{\mbox{\texttt{ZB20a}}}
\newcommand{\ZBXXb}{\mbox{\texttt{ZB20b}}}
\newcommand{\ZBXXtx}{\mbox{\texttt{ZB23x}}}

\newcommand{\LaXx}{\mbox{\texttt{La10x}}}

\newcommand{\LaXb}{\mbox{\texttt{La10b}}}
\newcommand{\LaXc}{\mbox{\texttt{La10c}}}
\newcommand{\LaXd}{\mbox{\texttt{La10d}}}
\newcommand{\Laiv}{\mbox{\texttt{La04}}}

\newcommand{\vb}[1]{\mbox{\boldmath$#1$}}
\newcommand{\nv}{\mbox{$\vb{n}$}}
\newcommand{\sv}{\mbox{$\vb{s}$}}
\newcommand{\vpi}{\mbox{$\varpi$}}
\newcommand{\Om}{\mbox{$\Omega$}}
\newcommand{\om}{\mbox{$\omega$}}
\newcommand{\OmE}{\mbox{$\Omega_E$}}
\newcommand{\pbar}{\mbox{$\bar{p}$}}
\newcommand{\ombar}{\mbox{$\bar{\omega}$}}
\newcommand{\Psibar}{\mbox{$\overline{\Psi}$}}
\newcommand{\PSI}{\mbox{$\Psi$}}
\newcommand{\Psin}{\mbox{$\Psi_0$}}
\newcommand{\pA}{\mbox{$p_A$}}
\newcommand{\alp}{\mbox{$\alpha$}}
\newcommand{\bet}{\mbox{$\beta$}}
\newcommand{\gam}{\mbox{$\gamma$}}
\newcommand{\ggp}{\mbox{$\gamma_{gp}$}}
\newcommand{\kap}{\mbox{$\kappa$}}
\newcommand{\x}{\times}
\newcommand{\e}[1]{\mbox{$\x10^{#1}$}}
\newcommand{\pmo}{\mbox{$^{-1}$}}
\newcommand{\beqn}{\begin{eqnarray}}
\newcommand{\eeqn}{\end{eqnarray}}
\newcommand{\nn}{\nonumber}
\newcommand{\q}{\frac}
\newcommand{\del}{\mbox{$\delta$}}
\newcommand{\D}{\mbox{$\Delta$}}

\newcommand{\tauD}{\mbox{$\tau$}}

\newcommand{\sm}{\mbox{$\sim$}}

\newcommand{\qq}{\qquad}

\newcommand{\asy}{\mbox{$''${\,}y\pmo}}
\newcommand{\lsim}{\raisebox{-.5ex}{$\stackrel{<}{\sim} \ $}}
\newcommand{\gsim}{\raisebox{-.5ex}{$\stackrel{>}{\sim} \ $}}
\newcommand{\Wms}{W~m$^{-2}$}
\renewcommand{\deg}{\mbox{$^\circ$}}

\newcommand{\gtfL}{\mbox{($g_2$$-$$g_5$)}}
\newcommand{\gftL}{\mbox{($g_4$$-$$g_3$)}}
\newcommand{\gfwL}{\mbox{($g_4$$-$$g_2$)}}
\newcommand{\gffL}{\mbox{($g_4$$-$$g_5$)}}
\newcommand{\sftL}{\mbox{($s_4$$-$$s_3$)}}
\newcommand{\stsL}{\mbox{($s_3$$-$$s_6$)}}
\newcommand{\sfsL}{\mbox{($s_4$$-$$s_6$)}}

\newcommand{\obl}{\mbox{$\epsilon$}}
\newcommand{\prc}{\mbox{$\phi$}}

\def\bls{1.0}
\def\blss{0.8}
\renewcommand{\baselinestretch}{\bls}

\journal{Earth-Science Reviews}

\graphicspath{{./}{figs/}}

\begin{document}

\begin{frontmatter}



\title{\bf Applying Astronomical Solutions and Milankovi{\'c} 
Forcing in the Earth Sciences}



\author{Richard E. Zeebe$^{1,*}$ and Ilja J. Kocken$^1$}
\address{\vspace*{0.5cm} 
     \normalfont
     $^*$Corresponding Author.\\ {\small
     $^1$School of Ocean and Earth Science and Technology, 
     University of Hawaii at Manoa, 
     1000 Pope Road, MSB, Honolulu, HI 96822, USA. 
     zeebe@soest.hawaii.edu, ikocken@hawaii.edu \\
     $[0000-0003-0806-8387]$ {Richard E. Zeebe} \\
     $[0000-0003-2196-8718]$ {Ilja J. Kocken}
     }       \\[2ex]
     Final revised version in press. \today  \\[2ex]
     
     \renewcommand{\baselinestretch}{1.0}\scs
     \tableofcontents
     \renewcommand{\baselinestretch}{\bls}\normalsize --- \\[10ex]
}

\renewcommand{\baselinestretch}{1.2}\normalsize
\begin{abstract}
Astronomical solutions provide calculated orbital and
rotational parameters of solar system bodies
based on the dynamics and physics of the solar 
system. Application of astronomical solutions in the Earth 
sciences has revolutionized our understanding in at least 
two areas of active research. ($i$)
The Astronomical (or Milankovi{\'c}) forcing of climate on 
time scales $\gsim$10~kyr
and ($ii$) the dating of geologic archives. The latter has permitted 
the development of the astronomical time scale, widely used 
today to reconstruct highly accurate geological dates and 
chronologies. The tasks of computing
vs.\ applying astronomical solutions are usually performed by 
investigators from different backgrounds,
which has led to confusion and recent inaccurate results
on the side of the applications.
Here we review astronomical solutions and Milankovi{\'c} forcing 
in the Earth sciences, primarily aiming at clarifying the astronomical 
basis, applicability, and limitations of the solutions.
We provide a summary of current up-to-date and outdated
astronomical solutions and their valid time span.
We discuss the fundamental limits imposed by dynamical 
solar system chaos on astronomical calculations and
geological/astrochronological applications.
We illustrate basic features of chaotic behavior using 
a simple mechanical system, i.e., the driven pendulum.
Regarding so-called astronomical ``metronomes'', we point out
that the current evidence does not support the notion of
generally stable and prominent metronomes for universal 
use in astrochronology and cyclostratigraphy.
We also describe amplitude and frequency modulation of
astronomical forcing signals and the relation 
to their expression in cyclostratigraphic sequences.
Furthermore, the various quantities and terminology
associated with Earth's axial precession are discussed 
in detail. Finally, we provide some suggestions regarding 
practical considerations.
\end{abstract}
\renewcommand{\baselinestretch}{\bls}\normalsize




\end{frontmatter}


%
%
\section{Introduction \label{sec:intro}}

\begin{figure}[t]
\vspace*{-50ex} \hspace*{+10.5ex}
\includegraphics[scale=0.65]{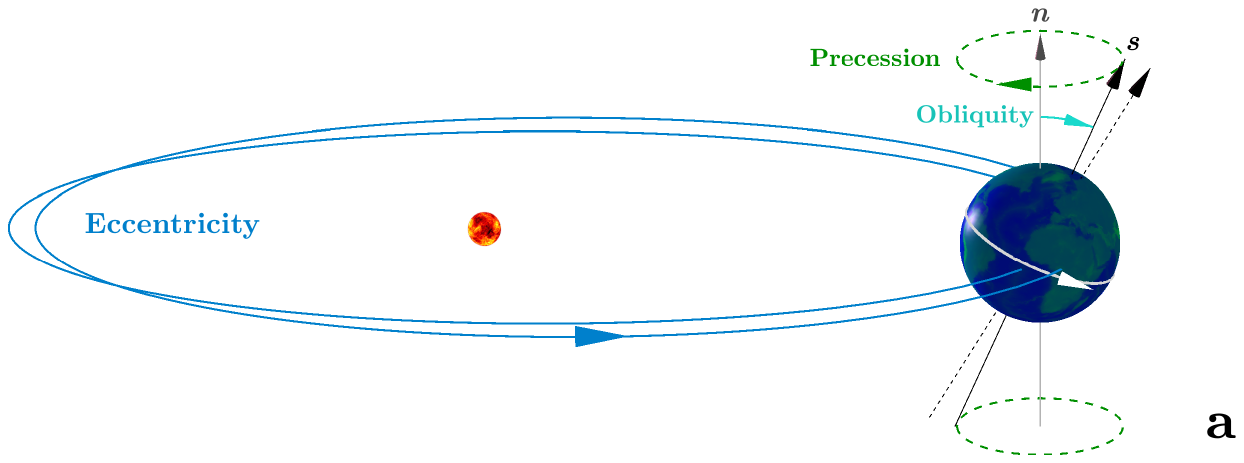}

\vspace*{-70ex} \hspace*{-00ex}
\includegraphics[scale=0.8]{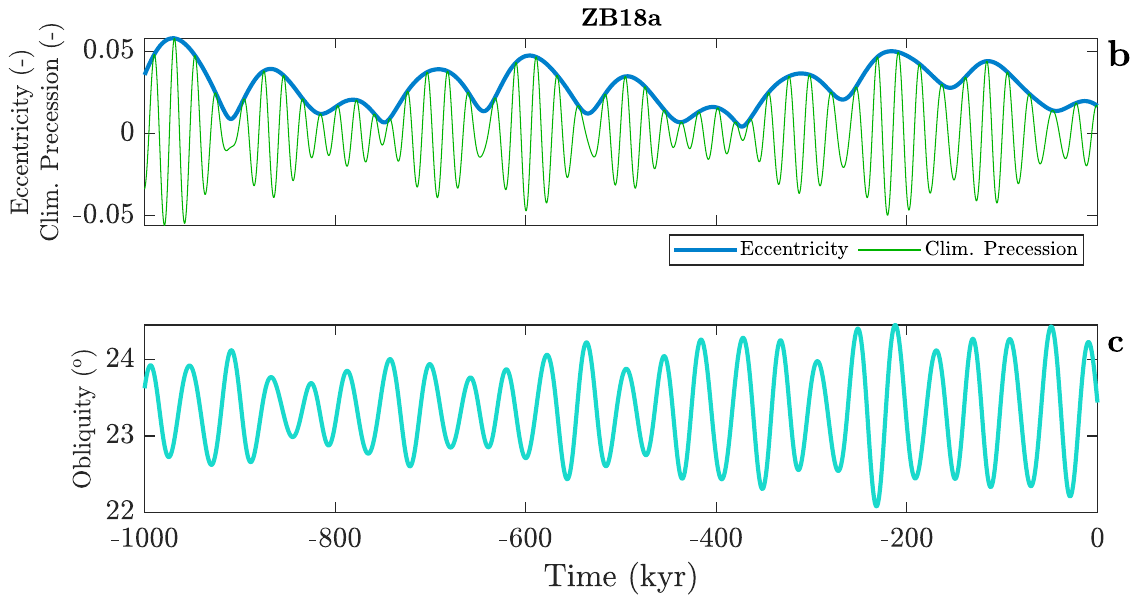}

\vspace*{-40ex}
\caption{\scs
Example of orbital parameters and their values frequently used 
in the Earth sciences.
(a) Illustration of orbital eccentricity (recent main periods 
\sm{100} and 405~kyr), precession (\sm{20}~kyr), and obliquity 
(\sm{41}~kyr), often called Milankovi{\'c} cycles (for geometrical 
illustrations of their physical meaning, see also 
Figs.~\ref{fig:eccInc} and~\ref{fig:illprc}).
For illustration, Earth's orbital eccentricity is highly
exaggerated. For an elliptical orbit, the sun is located at 
one of the ellipses' foci.
Obliquity is the angle between Earth's spin
axis \rev{(spin vector \vb{s})}
and the orbit normal (unit vector \vb{n} perpendicular to 
Earth's orbital plane = ecliptic, see Table~\ref{tab:notval}).
Viewed from ecliptic north, Earth's rotation (spin) is 
counterclockwise (from west-to-east, or eastward). 
Earth's orbital motion is in the same direction (prograde).
Axial precession (precession of the equinoxes) is in
the opposite direction (retrograde, green arrow).
Orbital parameter values shown here are from/based on 
the orbital solution \ZBETa\ \citep{zeebelourens19,
zeebelourens22pa} over the past 1~Myr.
(b) Earth's orbital eccentricity (blue) from \ZBETa\
and climatic precession (green) based on \ZBETa.
(c) Obliquity of Earth's spin axis based on \ZBETa.
\label{fig:eop}
}
\end{figure}

The term ``astronomical solution'' (\AS) as used here refers 
to calculated planetary orbital and rotational
parameters over time based on known
solar system physics at a specified point 
in time. Mathematically, an \AS\ for the planetary orbits
represents a solution 
to the equations of motion of the solar system with initial 
conditions at time $t_0$ for the positions, velocities, 
and masses of the solar system bodies included in
the calculation. Here we will focus specifically on \ASs\
that have direct applications in the Earth sciences, for 
instance, in geology,
paleoclimatology, astrochronology, and cyclostratigraphy; 
i.e., \ASs\ that provide accurate values for Earth's orbital 
parameters in the past (for a specific example, see Fig.~\ref{fig:eop}).
Thus, while a general review of \ASs\ may start historically with 
the work of Johannes Kepler, we start with
the work of Milutin Milankovi\'c, a Serbian engineer and
mathematician (1879-1958). 

Although not the first to propose 
astronomically-driven climate change on Earth 
\citep[for a review, see][]{emiliani95}, Milankovi{\'c} worked out an
astronomical theory to calculate the effects of secular
(`slow')
changes in insolation on Earth's climate and was primarily
interested in applying the astronomical theory to the ice age 
problem. In 1941, Milankovi{\'c} 
summarized his findings spanning thirty years of work in a 633-page 
volume to the Serbian Royal Academy \citep{milank41}. 
Milankovi{\'c} used an \AS\ that included elements of
Leverrier's work and contributions from Mi\v{s}kovi{\'c} 
\citep[see][p. 335 ff.]{milank41}.
Notably, Milankovi{\'c}'s astronomical theory of paleoclimate
was far from accepted at the time
and was virtually ignored for decades before being recognized by a 
larger community that realized its merit \citep[e.g.,][]{emiliani55Natb,
broecker66ast,imkip71Natb,hays76,berger78Natb}.
Today, there is overwhelming evidence that Earth's climate is 
paced by astronomical cycles on time scales $\gsim$10~kyr.
Based on work from the late 1800s, a new \AS\ was calculated 
by \citet{brouwer50} and subsequently applied to test the
astronomical theory of climate change
\citep[e.g.,][]{woerkom53,broecker66ast}.
Importantly, the \ASs\ up to that point were based on so-called
low-order perturbation theory, sometimes dubbed secular analytic 
theory, or Laplace-Lagrange solution \citep[for details and
history, see][]{murraydermott99,ito07}, which only allows 
for quasi-periodic, i.e., 
non-chaotic, solutions. A simple example of a quasi-periodic
function is the sum of two periodic functions (single
frequency each) with an irrational frequency ratio.

\def\cau{$1.495978707\e{11}$}
\def\cgm{$1.32712440041\e{20}$}
\def\kgg{$0.01720209895^2$}
\def\ugm{$\rm m^3~s^{-2}$}
\def\com{$7.292115\e{-5}$}
\def\crn{$3.8440\e{8}$}
\def\cen{$0.00327381$}
\def\cse{$328900.5596$}
\def\cel{$81.300568$}
\def\cgk{$0.9925194$}
\def\psn{$50.3848$}
\def\psnM{$7.597$}
\def\gdp{$-0.0192$}
\renewcommand{\baselinestretch}{\blss}
\begin{table}[t]
\caption{Notation and values employed in this paper (frequently
used variables).
\label{tab:notval}}
\begin{tabular}{lllll}
\hline
Symbol       & Meaning              & Value I/A & Unit  & Note  \\
\hline     
$a$          & Semimajor axis           &      & au    & Element 1 \\ 
$e$          & Orbital eccentricity     &      & --    & Element 2 \\
$I$          & Orbital inclination      &      & deg   & Element 3 \\
\Om          & Orbit LAN    \ $^a$      &      & rad   & Element 4 \\
\om          & Orbit AP     \ $^b$      &      & rad   & Element 5 \\
$\nu$        & True anomaly \           &      & rad   & Element 6 \\
$\vpi$       & Orbit LP     \ $^c$      &      & rad   & see text  \\
$N$          & Total No.\ of bodies \ $^d$ &   & --    &       \\
$\Sun, \ \Earth, \ \Moon$      
             & Sun, Earth, Moon/Lunar   &      & --    & or subscript $_{S,E,L}$ \\
\MS          & Central mass             &      & 1.0  or kg  & \\
$m_i$        & Mass of body $i$         &      & \MS\ or kg  & \\
$n_i$        & Mean motion              &      & rad~d\pmo   & \\
\obl         & Obliquity angle          &      & deg   &       \\
\obl$_0$     & Obliquity Earth $t_0$    & 23.4392911 & deg & \scs\citet{fraenz02}  \\
\prc         & Precession angle \ $^e$  &      &       &       \\
$\Psin$      & Luni-solar prec. rate $t_0$\ $^f$ & \psn & \asy  & \scs\citet{capitaine03}     \\
$\Psi_A$     & Accumulated angle        &      &  $''$ & $\dot{\Psi}_A = \Psi$     \\
$p_A$        & General prec. in long.$^g$      &       & $''$  &  $p_A = - \prc$   \\
\ombar       & Orbit LPX $^h$           &      &       & $\ombar=\vpi+|\prc|$ \ $^e$  \\
\pbar        & Climatic precession      &      &       & $\pbar = e \sin \ombar$ \\ 
\sv          & Spin vector              &      &       &       \\ 
\nv          & Orbit normal             &      &       &       \\ 
$H$          & Dynamical ellipticity    & \sm{0.00328} & --    &       \\ 
$k^2$        & (Gauss grav. const.)$^2$ & \kgg & see $^i$  &   \\
au           & Astronomical unit        & \cau & m     &       \\
$G\MS$       & Sun GP $^j$              & \cgm & \ugm  &       \\ 
$\MS/(\mE+\mL)$ & Mass ratio            & \cse & --    &       \\
$\mE/\mL$    & Mass ratio               & \cel & --    &       \\
\OmE         & Earth's angular speed    & \com & rad~s\pmo & at $t_0$ \\
\hline
\end{tabular}

{\scs
$^a$ LAN = Longitude of Ascending Node.
$^b$ AP  = Argument of Perihelion.
$^c$ LP  = Longitude of Perihelion.
$^d$ Including the central mass.
$^e$ Axial precession is retrograde (Section~\ref{sec:lscp}, 
     Fig.~\ref{fig:illprc}), hence $\dot{\prc}_0$ is taken negative here
     (time derivatives are denoted by ``dot'', $\dot{\prc} = d\phi/dt$).
$^f$ Conventionally, $\Psi$ is taken positive (Section~\ref{sec:lscp}).
$^g$ General precession in longitude.
$^h$ LPX = LP from the moving equinox.
     \ombar\ (``omega bar'') is not be confused with \vpi\ (``varpi'').
     Various symbols are used for \ombar\ in the literature.
$^i$ General unit of $k$: $\rm au^{3/2}~d\pmo~M^{-1/2}_0$,
     dimensionless in astronomical units, also rad~d\pmo\ when 
     e.g., equated with $n_i$.
$^j$ GP = Gravitational Parameter.
}
\end{table}
\renewcommand{\baselinestretch}{\bls}

From a practical perspective, the next significant step in 
\AS\ development followed in the 1970s and provided Earth's 
orbital parameters over the past few million years at
improved accuracy \citep{bretagnon74,berger76ast,berger77}.
The \ASs\ were still based on secular perturbation theory
but included higher-order terms in series
expansions with respect to eccentricity, inclination, 
and planetary masses \citep{bretagnon74,duriez77,
laskar85}. Taking advantage of the accelerating computer 
power in the 1980s, \citet{sussman88} studied the 
outer planets using full numerical integrations over 845~Myr
\rev{on the Digital Orrery}
and showed that the motion of Pluto is chaotic.
\rev{(The Digital Orrery was a special-purpose computer built 
specifically for studying planetary motion \citep{applegate85},
named after {\it orrery}, a mechanical model of the solar system
that illustrates the relative positions and motions 
of the planets and moons.)}
Higher-order, averaged equations of the secular
evolution of the inner and outer planets
also showed chaos, with a ``Lyapunov time'' for the inner solar 
system of only $\sim$5~Myr \citep{laskar89}.\footnote{
``Myr'' (million years) is used here for duration, length of time 
intervals, and numerical time in \ASs\ (negative in the 
past), whereas ``Ma'' (mega annum) is used for geohistorical 
dates; correspondingly for kyr, ka, Gyr and Ga
\citep{aubry09}.
}
The Lyapunov time
is a metric to characterize chaos in dynamical systems and
represents the time scale of exponential divergence (e-folding 
time) of nearby trajectories \citep[see, e.g.,][]{murraydermott99}.

The first fully numerical and ``direct'' long-term integration 
(the past 3~Myr) of the eight planets and Pluto {\sl directly} based 
on the equations of motion (not analytical)
was carried out by \citet{quinn91} using a multistep method.
\citet{quinn91} showed that the model and secular theory
predictions of, e.g., Earth's orbital eccentricity and obliquity 
by \citet{berger78Natb}, were inaccurate beyond about 1-1.5~Myr in the 
past, although subsequent updates by \citet{bergerloutre92} 
showed better agreement with \citet{quinn91}. 
Notably, the \ASs\ based on earlier physics models and
low/intermediate-order 
analytic secular perturbation theory were (a) by design unable 
to reveal the chaotic behavior of the the solar system and (b) 
were only accurate over the past few Myr. These results
suggested that either fully numerical approaches or 
very high-order secular solutions were required for accurate 
applications in the Earth sciences. \citet{laskar90} provided a long-term
orbital solution that was obtained using an
extended, averaged secular system.
Regarding long-term solar system integrations, we note that
\citet{ito02} performed full numerical, long-term integrations
of up to 5~Gyr of the eight planets and Pluto.
However, the study's focus was long-term stability, rather 
than accuracy for geological applications and omitted
several second-order effects (i.e., general relativity,
a separate Moon, and asteroids).
In 2003, Varadi and co-workers published the results of a 
direct, fully numerical integration 
of the eight planets over the past 100~Myr using a St\"ormer multistep 
scheme, which revealed large differences to the results of \citet{laskar90} 
already around $-24$~Myr \citep{varadi03}
\rev{(past time in \ASs\ is negative here, see 
footnote~1)}. The early 2000s effectively
marked the end of \AS\ based on secular perturbation theory for 
any serious geological application. Subsequent \AS\ employed in,
for instance, astrochronology and cyclostratigraphy
are based on full numerical integrations \citep{laskar04Natb,laskar11,
zeebe17aj,zeebelourens19,zeebelourens22pa,zeebelourens22epsl,
zeebelantink24aj,zeebelantink24pa}.
The most recent numerical solutions are described in more detail 
in Section~\ref{sec:u2dOS}.

Over the past few decades, the improvement in accuracy of \ASs\ 
has permitted the development of the astronomical time scale 
\rev{(ATS, see Table~\ref{tab:aa})}, 
which has transformed the dating of geologic archives.
Simply put, the ATS represents an accurate astronomical calendar 
used in the Earth sciences based on the motion of solar system bodies
to study and explain Earth's geologic history
\citep[for further reading, see, e.g.,][]{montenari18}.
In addition to accurate relative (floating) age models and
chronologies \rev{(see also Section~\ref{sec:deep})},
astrochronology provides highly accurate geological
dates with small error margins. For example, recent efforts
have dated the Paleocene-Eocene Thermal Maximum (PETM) onset
at $56.01 \pm 0.05$~Ma and the Cretaceous-Tertiary Boundary
at $65.96$ to $65.52$~Ma
\citep{zeebelourens19,zeebelourens22epsl}. ``Tertiary'' is 
used informally here, not as a formal division \citep{ics05}.
Astrochronology has made formidable progress in geological dating 
through deep-sea drilling and age-model
tuning to \ASs, which led to the
calibration of critical intervals of the geologic time scale, 
particularly in the early and mid-Cenozoic
\citep[e.g.,][]{hinnov00,zachos01Natb,lourens05,westerhold08,
hilgen10,hilgen15,
liebrand16,meyers18,lauretano18,likump18,hinnov18,zeebelourens19,
westerhold20,zeebelourens22pa,zeebelourens22epsl}.

\rev{
Importantly, an increasing number of studies at the current frontier 
of cyclostratigraphic research are pushing the boundaries into 
deep time, thereby providing unique insight into astronomical
forcing of paleoclimate and solar-system evolution
over hundreds of millions to billions of years
\citep[e.g.,][]{zhang15orb,mameyers17,meyers18,kent18,
lantink19,olsen19,soerensen20,lantink22,lantink23,lantink24,
zeebelantink24aj,zeebelantink24pa,malinverno24,wumalinverno24}. 
A critical ongoing task is to use the 
geological record to confirm and map the solar system's chaotic 
behavior, reconstruct the Earth-Moon history, and develop a 
``Geological Orrery'', in analogy to the mechanical and Digital 
Orrery \citep[see above and][]{olsen19}. For more information
on astronomical forcing and astronomical solutions in deep time, 
see Section~\ref{sec:deep}.
}

\renewcommand{\baselinestretch}{\bls}\selectfont
\begin{table}[t]
\rev{
\caption{Frequently used acronyms
\label{tab:aa}}
\vspace*{5mm}
\hspace*{-5mm}
\begin{tabular}{ll}
\hline
Acronym & Meaning \\
\hline \\[0ex] 
\AS    &  Astronomical Solution   \\
AM     &  Amplitude Modulation    \\
ATS    &  Astronomical Time Scale \\
EC     &  Eccentricity Cycle             \\
ETP    &  Eccentricity, Tilt, Precession \\
FM     &  Frequency Modulation           \\
FFT    &  Fast Fourier Transform         \\
LEC    &  Long Eccentricity Cycle        \\
OS     &  Orbital Solution               \\
PT     &  Precession-Tilt                \\
SEC    &  Short Eccentricity Cycle       \\
VLEC   &  Very Long Eccentricity Cycle   \\
VLN    &  Very Long eccentricity Node    \\
\hline
\end{tabular}
\noindent {\scs \\[2ex]
}
}
\end{table}
\renewcommand{\baselinestretch}{\bls}\selectfont

Today, \ASs\ and the ATS represent the backbone of astrochronology 
and cyclostratigraphy, and are widely used in the Earth sciences,
including areas of geology, geophysics, paleoclimatology, 
paleontology, and more. The applications are broad, ranging from
high-fidelity dating (establishing highly accurate geological ages
and chronologies) and reconstructing forcing/insolation patterns and 
their effects on paleoclimate --- to the evolution of the Earth-Moon system
and nonlinear dynamics \citep[to name just a few, for recent 
summaries, see][]{montenari18,hinnov18,cvijanovic20,lourens21,
devleesch24,wumalinverno24}.

\subsection{The Main Milankovi{\'c} Cycles \label{sec:milank}}

The focus of this review is on astronomical solutions and
forcing, rather than on the forcing's effects on Earth's 
climate through insolation changes. Nevertheless, the main 
Milankovi{\'c} cycles are
very briefly summarized here \citep[for details and reviews, see, 
e.g.,][]{milank41,hays76,broecker85habNatb,berger94,muller02,
hinnov18,lourens21}. As mentioned above, the main 
Milankovi{\'c} cycles usually refer to precession, obliquity, and 
eccentricity (for illustration and main periods, see 
Fig.~\ref{fig:eop}). To be precise, we refer here to climatic 
precession (see Section~\ref{sec:lscp}). 
Climatic precession and obliquity affect the
geographical distribution of insolation over time, whereas
eccentricity affects the total insolation Earth receives
(precession and obliquity do not).
For instance, climatic precession shifts the positions
of the equinoxes relative to Earth's eccentric orbit
(see Figs.~\ref{fig:eop} and~\ref{fig:illprc}). Hence,
if at a given time northern hemisphere summer occurs 
at perihelion, it will occur at aphelion about 10~kyr 
later (perihelion and aphelion refer to the closest and 
farthest distance from the Sun along Earth's orbit, see 
Fig.~\ref{fig:eccInc}). As a result, the insolation
at a given calendar day and latitude on Earth varies
with climatic precession over time \citep[one critical element 
in the pacemaking of the ice ages, e.g.,][]{hays76}.
The effects of climatic precession are anti-phased between
the hemispheres.

Obliquity changes the solar incidence angle of insolation 
at a given latitude on Earth (see Figs.~\ref{fig:eop} 
and~\ref{fig:illprc}).
Obliquity thus affects the seasonal contrast of insolation
(in-phase between the hemispheres and increasingly pronounced
toward higher latitudes).
For instance, at an obliquity angle $\obl = 0\deg$,
the tilt-induced seasons would disappear, whereas at $\obl = 
90\deg$, the tilt-induced seasons would be extreme 
(as is the case on Uranus with an obliquity of \sm{98}\deg).
Earth's recent obliquity varies between minima and maxima
of about $22.0\deg$ and $24.5\deg$ over a 41-kyr main period
(Fig.~\ref{fig:eop}), which causes substantial
changes in Earth's climate --- often enhanced when continental
ice sheets are present \citep[for summaries, see, e.g.,][]
{lourens21,devleesch24}.
Local insolation changes due to
obliquity (and climatic precession) may be sizable.
For example, at $65\deg$N (Jun 21) insolation varies up to 
\sm{120}~\Wms\ over the past 1~Myr.

Earth's orbital eccentricity (\eE) impacts insolation
in multiple ways. For example, on precessional time
scales, \eE\ modulates the amplitude of precession,
which, in turn, affects insolation
(Fig.~\ref{fig:eop}). On annual time scales, \eE\
influences the total insolation 
Earth receives along its orbit, as the distance ($r$) to the 
Sun varies (see Fig.~\ref{fig:eop} and~\ref{fig:eccInc},
for $\eE \neq 0$). At perihelion and aphelion $r$ is
equal to $a(1 - \eE)$ and $a(1 + \eE)$, respectively,
where $a$ is the semimajor axis \citep[e.g.,][]{danby88}. 
Given that the insolation with respect to Earth's cross section
at distance $r$ is proportional to $r^{-2}$, the insolation 
ratio at perihelion vs.\ aphelion is:
\beqn
(1 + \eE)^2 / (1 - \eE)^2 \ ,
\eeqn
which yields \sm{7\%} and 27\% at present $\eE = 0.0167$ 
and $\max\{\eE\} \simeq 0.06$, respectively (see 
Fig.~\ref{fig:eop}). As a result, eccentricity also 
induces ``seasons'', which are, however, hemispherically 
symmetric and presently smaller than the tilt-induced seasons.
The {\it total mean annual insolation} (or energy $W$)
Earth receives is proportional to 
\citep[e.g.,][]{berger94}:
\beqn
W \propto \left( 1 - \eE^2 \right)^{-\q{1}{2}} \ .
\label{eqn:insol}
\eeqn
The absolute effect of the factor $\left( 1 - 
\eE^2 \right)^{-\q{1}{2}}$ at present is small 
(\sm{0.014\%}), compared to a circular orbit with 
$\eE = 0$. However, for long-term 
variations, the ratio of $W$ at different \eE's matters. 
For example, between the present $\eE = 0.0167$  and 
$\eE = 0.06$, $W$ increases by \sm{0.17\%},
which is 
{\sl not} negligible because it raises the total, global 
energy received. At a solar constant of 1370~\Wms, 0.17\%
amounts to 2.3~\Wms\ (0.6~\Wms\ when distributed over 
Earth's surface, i.e., reduced by factor 4). For 
comparison, a doubling of CO$_2$ in Earth's atmosphere is 
equivalent to \sm{3.7}~\Wms\ of radiative forcing.
Thus, orbital eccentricity forcing (in \Wms) may appear small 
compared to local changes from climatic precession and 
obliquity (see above). However, the difference is that
eccentricity forcing is global.
Finally, and regardless of whether precession, obliquity,
or eccentricity forcing is involved, caution is advised 
when attempting to predict the impact of astronomical 
forcing on Earth's climate, because the climate system 
response is highly non-linear.

\rev{
Regarding the stratigraphic recording of the climate
system response (say, in sedimentary sequences), it 
is noteworthy that the response of 
the sedimentary system to climate perturbations is also 
inherently non-linear. Thus, the translation of astronomical 
cycles to sedimentary cycles is modified by two non-linear 
transfer functions.
}

\subsection{Types of Astronomical Solutions}

Two types of astronomical solutions are discussed in the following:
orbital solutions (OSs) and precession-tilt (PT) solutions.
OSs describe the orbital dynamics of the solar system, usually
considering the solar system bodies as mass points (as is the case 
here). OSs provide values for the orbital elements of solar system 
bodies over time, including orbital eccentricity and inclination 
(see Fig.~\ref{fig:eccInc}). PT solutions 
describe the rotational dynamics of individual solar system bodies (here 
of the Earth), considering the physical dimensions and shape of the
body. PT solutions provide values for precession and obliquity 
over time, e.g., based on spin axis dynamics (see Fig.~\ref{fig:illprc}).
OS dynamics have important effects on (and are a prerequisite for) 
PT solutions, while the effect of rotational dynamics on OSs
is generally minor \citep[e.g., Earth-Moon dynamics, see][]
{zeebelantink24pa}. For example, amplitude 
variations in Earth's orbital inclination (due to OS dynamics) 
are reflected in obliquity --- most evidently during intervals 
of reduced amplitude variation \citep[see][]{zeebe22aj}. 
Similarly, amplitude variations in eccentricity (due to OS 
dynamics) are reflected in climatic precession. 

\subsection{Organization of Content}

The remainder of this review is organized into three sections
focusing on orbital solutions, precession-tilt solutions, and 
practical considerations (Sections~\ref{sec:os}, \ref{sec:pt},
and \ref{sec:pract}). Section~\ref{sec:os} introduces several
basic concepts for describing and analyzing orbital solutions,
including orbital elements and the fundamental frequencies
of the solar system. Solar system chaos is key and indispensable 
to understanding a variety of topics, including
the limitations imposed on orbital solutions by chaotic
dynamics. Chaos
is therefore discussed in some detail under orbital 
solutions (Section~\ref{sec:chaos}).
Chaos also causes critical changes in the amplitude modulation (AM)
of orbital forcing signals, which, if expressed in cyclostratigraphic
sequences, can be used to reconstruct the solar
system's chaotic history. Section~\ref{sec:chaos} 
on solar system chaos thus precedes Section~\ref{sec:amfm} 
on amplitude and frequency modulation. Furthermore,
AM originates from combination 
terms related to the fundamental frequencies but
can also be studied using eccentricity and obliquity analysis.
Thus, AM is examined in some depth in two sections 
(\ref{sec:fcomb} and~\ref{sec:am}). Up-to-date orbital
solutions, their valid time span (another consequence of solar 
system chaos), and the confusion surrounding the subject
are described in Section~\ref{sec:u2dOS}. Importantly,
the probability that a particular OS represents 
the actual, unique history of the solar system is near 
zero significantly beyond the OS's valid time span
\rev{(the details and implications are explained in 
Sections~\ref{sec:u2dOS}
and~\ref{sec:pract})}. Section~\ref{sec:pt}
dives into the details of precession-tilt solutions, explaining
the various quantities associated with Earth's axial precession
and aiming at clarifying the terminology and notation used in the 
literature. Up-to-date precession-tilt solutions are described 
in Section~\ref{sec:u2dPT}, as well as recently 
employed tools that produce inaccurate results.
\rev{Another critical application of astronomical
solutions beyond the direct use for astronomical tuning 
(e.g., absolute Cenozoic ages and chronologies) is
astronomical forcing and astrochronology in deep time 
(see Section~\ref{sec:deep}).
}
Section~\ref{sec:pract} provides several recommendations
for the user seeking guidance and lists selected resources for 
practical consideration. We close with a brief summary and 
outlook (Section~\ref{sec:summ}).

%
\section{Orbital Solutions \label{sec:os}}

\subsection{Fundamentals}

\subsubsection{Orbital Elements}

\begin{figure}[t]
\vspace*{-35ex} \hspace*{-05ex}
\includegraphics[scale=0.8]{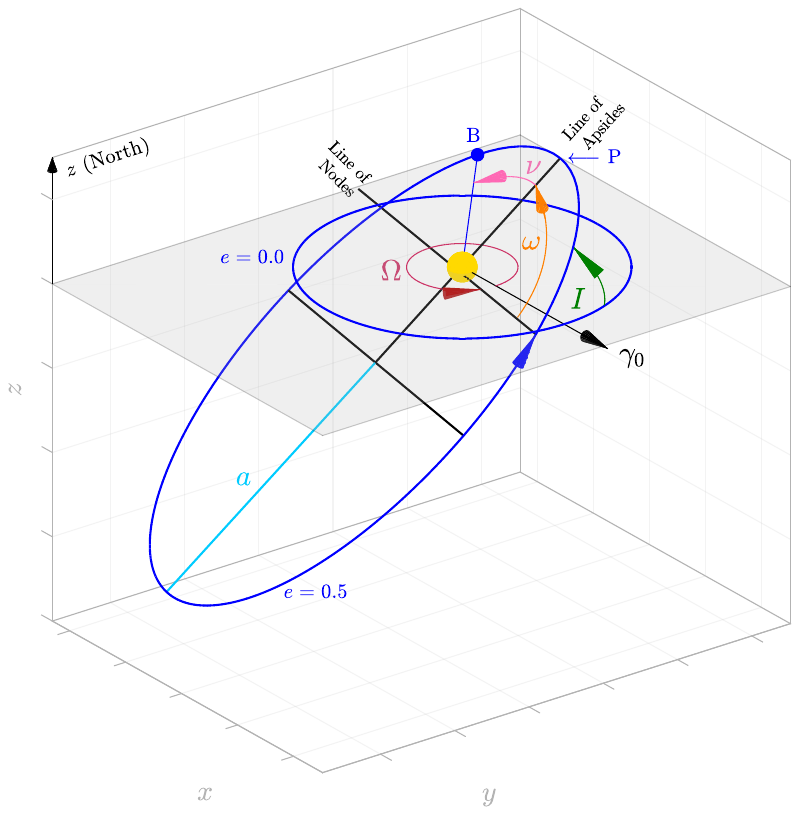}

\vspace*{-30ex}
\caption{\scs
Illustration of elements characterizing 
a Keplerian orbit. $a = $ semimajor axis, $e = $ 
eccentricity, $I = $ inclination. $I$ is the angle measured 
from the fixed reference plane (gray) to the instantaneous orbit plane. 
Two orbits (blue, different $a$'s) are shown; the circular 
orbit ($e = 0$, $I = 0$) 
lies in the reference plane; the elliptic orbit ($e = 0.5$) is inclined
relative to the reference plane.
$\Om = $ longitude of the ascending node is measured in the reference 
plane from the reference point $\gam_0$ to the line of nodes
where the orbiting body (B) ascends, i.e., passes from south to north 
($z < 0$ to $z > 0$, see blue arrow) through the reference plane.
Hence \Om\ (if defined) is measured in a plane different from
B's orbital plane.
In the solar system, $\gam_0$ may refer to the vernal point
at a fixed date.
$\om = $ argument of 
perihelion is measured from the ascending node to the 
perihelion (P).
The line of apsides passes through the perihelion (P) and 
aphelion of the orbit (closest and farthest distance from 
the central body).
$\nu = $ true anomaly is measured from the perihelion to
the orbiting body (B).
\label{fig:eccInc}
}
\end{figure}

For the reader unfamiliar with orbital parameters commonly 
used in astronomy, we first introduce the Keplerian elements
and the notation used throughout this paper 
(see Fig.~\ref{fig:eccInc} and Table~\ref{tab:notval}). Keplerian 
orbits include circular ($e = 0$), elliptic ($0 < e < 1$,
$a > 0$), 
parabolic ($e = 1$), and hyperbolic ($e > 1$,  $a < 0$)
orbits. To describe the orbit, six independent elements are used,
five of which (for instance, $a, e, I, \Om$, and $\om$) define the 
shape and orientation of the orbit in space (for illustration,
see Fig.~\ref{fig:eccInc}). The final element (such as the true 
anomaly $\nu$) identifies the body's position in the orbit
and hence represents the instantaneous (rapidly changing)
orbital element. Notably, the Keplerian elements are used to 
describe orbits in general, not only exact solutions to Kepler's 
two-body problem, which can be solved analytically. In the two-body
problem ($N = 2$), $\nu$ varies over time, while the five elements 
defining orbit shape and orientation are constant, which is not the
case in general ($N > 2$). For example, the motion of solar system bodies 
such as the planets include the full planet-planet interactions 
and hence do not evolve on simple Kepler ellipses. Nevertheless,
at a given instant in time, the orbit can be defined by Keplerian 
(also called osculating) elements, which refer to the orbit the body
would have without perturbations (in mathematics osculate means to touch 
so as to have a common tangent). In the general case, the orbit's shape
and orientation change over time. The six orbital elements can
be converted into the body's Cartesian position and velocity
vectors \vb{x} and \vb{v} (and vice versa)
via a coordinate transformation (six degrees of freedom).
Thus, the output of orbital solutions may be provided
in orbital elements or Cartesian vectors, or both.

Note that for $I = 0$, the line of nodes and hence \Om\ and \om\
are undefined. Nevertheless, as long as $e \neq 0$, the longitude
of perihelion, \vpi, may always be defined \citep[see Section 2.3
in][]{bate71}. \vpi\ is measured from the reference point to 
the perihelion eastward (to the ascending node, if it exists, and 
then in the orbital plane to the perihelion). If both \Om\ and \om\ 
are defined, then
\beqn
\vpi = \Om + \om \ .
\eeqn
In general, \vpi\ is therefore a ``dogleg'' angle, as \Om\ and \om\
lie in different planes. However, this does not impede its use
and utility as an orbital element. Importantly, \vpi\ is well 
defined for all inclinations.

\subsubsection{Eccentricity and Inclination \label{sec:einc}}

\begin{figure}[t]
\vspace*{-55ex} \hspace*{-07ex}
\includegraphics[scale=0.9]{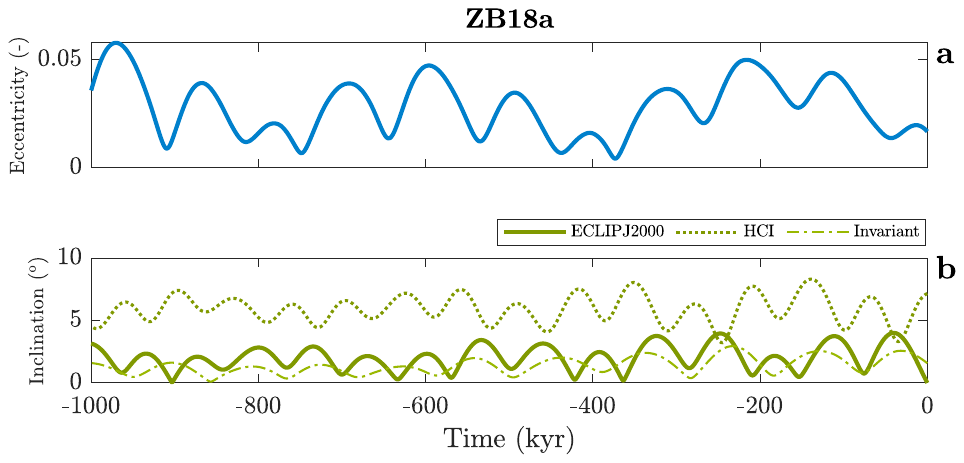}

\vspace*{-50ex}
\caption{\scs
Earth's orbital eccentricity and inclination from \AS\ \ZBETa\
\citep{zeebelourens19} over the past 1~Myr.
For geometrical illustrations of the physical meaning of 
orbital eccentricity and inclination, 
see Figs.~\ref{fig:eccInc}, \ref{fig:gsmod}, and~\ref{fig:illprc}.
(a) Earth's orbital eccentricity (which is independent of
the reference frame).
(b) Earth's orbital inclination in different reference
frames. Solid line: Ecliptic frame (ECLIPJ2000), dashed
line: Heliocentric Inertial (HCI) frame
\citep[for details, see][]{fraenz02,zeebe17aj},
dot-dashed line: invariant frame (see text).
\label{fig:ei}
}
\end{figure}

Earth's orbital eccentricity is among the most frequently used orbital 
parameters in Earth-science applications (see Fig.~\ref{fig:ei} 
for an example over the past 1~Myr). Cyclostratigraphic sequences, 
for instance, 
often show strong cyclicity at eccentricity frequencies
(main periods of \sm{100}~kyr and 405~kyr in the recent past). 
Eccentricity affects the total insolation Earth receives 
over one year \citep[e.g.,][]
{berger94}, which in turn affects Earth's climate
(Section~\ref{sec:milank}). 
Moreover, if clear axial precession signals are present 
in the geologic
sequence, the observed precession amplitude is modulated
by eccentricity (in a very simple manner, i.e., 
eccentricity is the envelope of climatic 
precession, see Fig.~\ref{fig:eop}). Eccentricity is related
to the shape of the orbit (not the orientation in space,
see Fig.~\ref{fig:eccInc}) and is hence independent of the
reference frame used to describe the orbit. In contrast,
orbital inclination depends on the reference frame,
as the choice of the reference plane is arbitrary
(Fig.~\ref{fig:eccInc}). Thus, the representation of
inclination may differ fundamentally between different
reference frames (for a frequently used frame such as
the ecliptic, see Fig.~\ref{fig:ei}). However, other 
frames such as the invariant frame (based on the invariable 
plane, perpendicular to the solar system's total angular 
momentum vector that passes through its barycentre)
or the Heliocentric Inertial (HCI) frame 
(based on the orientation of the solar equator
at a fixed point in time) are 
equally valid and are useful for e.g., accounting for effects 
of the solar quadrupole moment on the dynamics 
\citep[for details, see][]{fraenz02,souami12,zeebe17aj}. 
Note that accurate 
transformations between frames generally require 
detailed information about the coordinate systems and, 
depending on the case, parameters such as initial conditions,
masses, constants used, etc. (see below and \citet{souami12}).

Because of the distinct 
physical nature of inclination and eccentricity
(see Fig.~\ref{fig:eccInc}), orbital forcing effects
(Section~\ref{sec:milank})
due to inclination and eccentricity are principally
different \citep[for further discussion, see, e.g.,][]
{muller02,zeebe22aj,zeebelantink24pa}.
Orbital inclination represents one of the controls on the 
obliquity of Earth's spin axis but the relationship is more 
complex than between eccentricity and climatic precession
($\pbar = e \sin \ombar$). 
For example, the obliquity amplitude (variation around 
the mean) depends on the main inclination amplitude 
and frequency but also on the luni-solar precession rate
\citep[see Section~\ref{sec:pt} and e.g.,][]{ward74,ward82,zeebe22aj,
zeebelantink24pa}.
In addition to their role as individual orbital parameters,
orbital inclination and eccentricity are key to understanding
the fundamental (secular) frequencies of the solar system
(Section~\ref{sec:gs}).

Different eccentricity cycles (ECs) may be expressed
and hence be observable 
in cyclostratigraphic sequences, i.e., the short~EC (SEC),
long~EC (LEC) and very~long~EC (VLEC). The forcing periods 
of the two main SEC pairs in the recent past are about 
95/99~kyr and 124/131~kyr, while the dominant LEC forcing period 
in the recent past is \sm{405}~kyr (see Section~\ref{sec:fcomb}). 
Until recently, the
widely accepted and long-held view was that the LEC was 
practically stable in the past and has been suggested
for use as a ``metronome'' (see Section~\ref{sec:mtrn})
to reconstruct accurate ages and 
chronologies, including deep-time geological applications
\citep{laskar04Natb,kent18,spalding18,meyers18,montenari18,
lantink19,devleesch24}. However, it has recently been 
demonstrated that the LEC can become unstable over long 
time scales \citep{zeebelantink24aj}. The VLEC refers
to the long-term amplitude modulation of eccentricity
(recent period of \sm{2.4}~Myr) and is related to the 
secular frequency term \gftL\ (see Section~\ref{sec:gs}).
The VLEC is unstable and its expression in 
cyclostratigraphic sequences may be used to reconstruct 
the solar system's chaotic history beyond \sm{50}~Ma 
\citep{mameyers17,westerhold17,olsen19,zeebelourens19}.
For more information on the unstable VLEC and
the resonance involving \gftL, see Section~\ref{sec:am}.

\subsubsection{Secular Frequencies: $g$ and $s$ Modes
\label{sec:gs}}

As mentioned above, in planetary systems with $N > 2$, the 
shape and orientation of the orbits change over time
due to mutual interactions. For example, the eccentricity and 
inclination of each orbit generally varies and the line of apsides 
and the line of nodes are not fixed in space as in the two-body problem
(see Fig.~\ref{fig:eccInc}). The lines generally precess, aka 
apsidal and nodal precession, respectively. 
Given that the Newtonian interaction between the orbiting 
bodies depends on mass and distance, one would expect that the 
frequencies at which the orbital elements change over time would
depend on, for instance, $m_i$ and $a_i$. Indeed, an analytical
calculation to determine, for example, the frequencies for
$N = 3$ (including one dominant mass)
using low-order secular perturbation theory for small
$e_i$ and $I_i$ shows a frequency dependence on only $m_i$ and 
$a_i$, and $n_i$, where $n_i = (k^2 \MS/a^3_i)^\q{1}{2}$
\citep{murraydermott99}. The frequency spectrum of each 
body is composed of different contributions from the
fundamental (secular, slowly changing) frequencies
of the full system (aka fundamental proper modes, or 
eigenmodes). The secular frequencies may be thought of as the
spectral building blocks of the system as given (dominated
by the $m$'s and $a$'s of the major bodies, e.g., the planets
in the solar system). Importantly, 
however, there is generally no simple one-to-one relationship 
between, say, an eigenmode and a single planet, 
particularly for the inner planets (note that there is no apsidal 
and nodal precession for a single body). The system's motion is 
a superposition of all eigenmodes, although some modes represent 
the single dominant term for some (mostly outer) planets. 
The breakdown into eigenmodes is similar to the
problem of $N$ coupled oscillators in physics, where the overall 
motion/spectrum may be complex but can be decomposed into the 
superposition of $N$ characteristic, fundamental frequencies.
In celestial mechanics, the lowest-order solution 
(in $e$ and $I$) to the $N$-body 
problem (with a dominant central mass and small $e_i$ and $I_i$) is 
called Laplace-Lagrange solution \citep[e.g.,][]{murraydermott99}.

\begin{figure}[t]
\vspace*{-26ex}\hspace*{+10ex}
\includegraphics[scale=0.55]{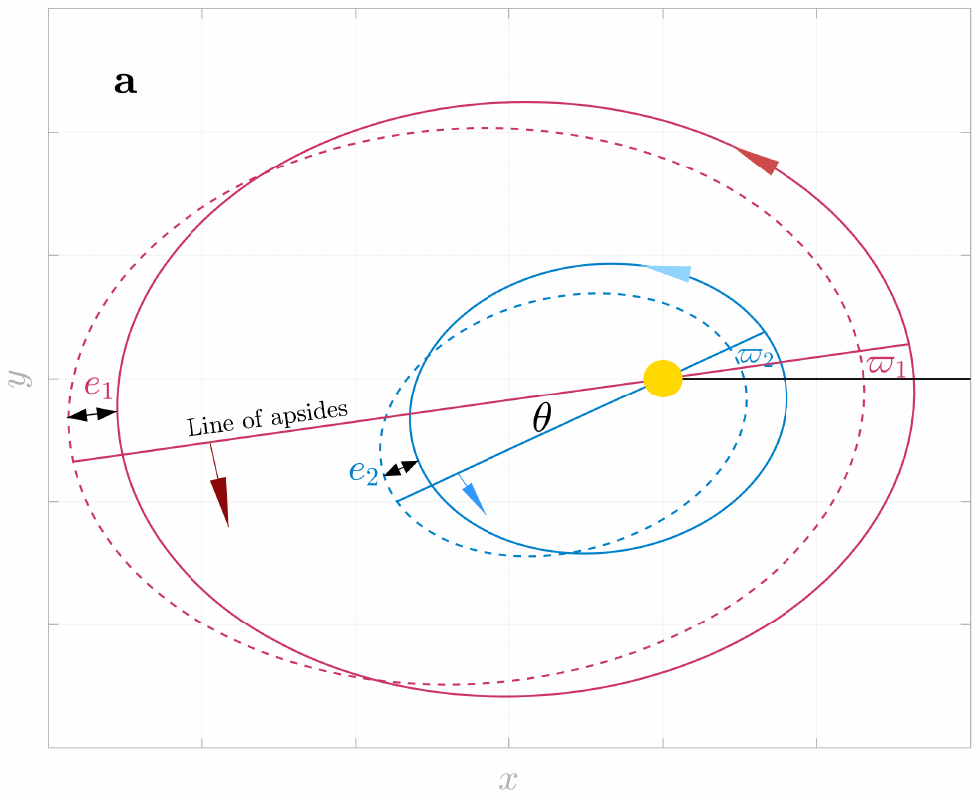}

\vspace*{-41ex}\hspace*{+10ex}
\includegraphics[scale=0.55]{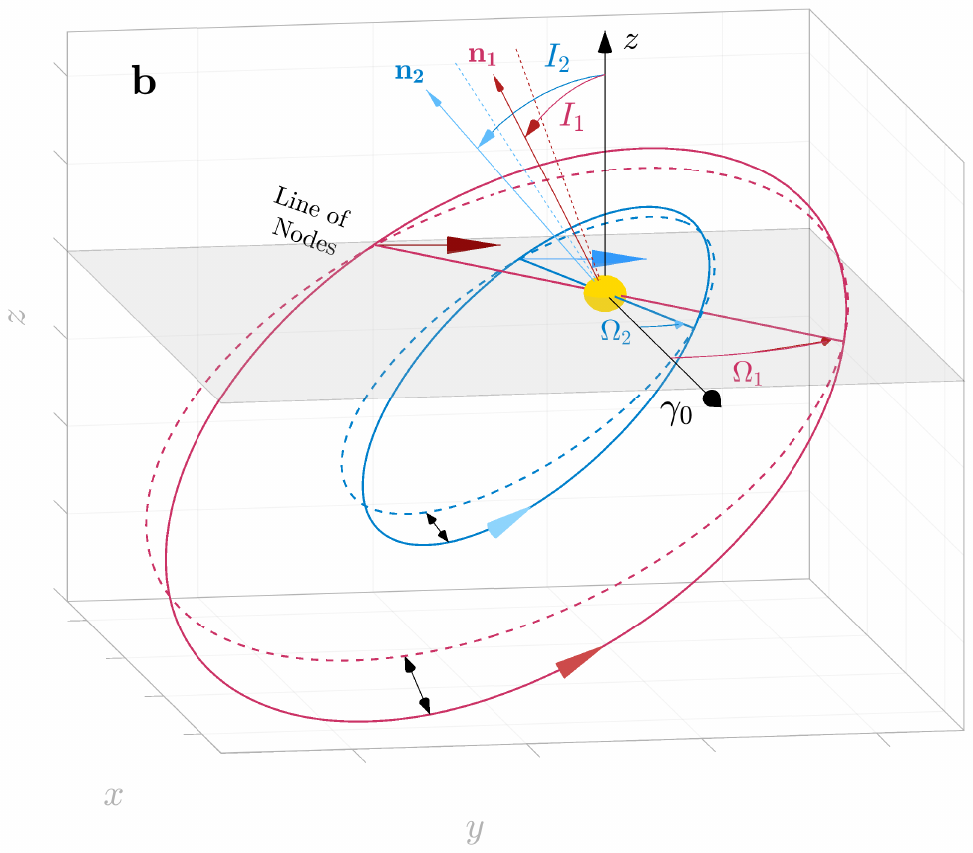}

\vspace*{-20ex}
\caption{\scs
Schematic illustration of $g$ and $s$ modes for $N = 3$
(for symbols and notation, see Table~\ref{tab:notval}
and Fig.~\ref{fig:eccInc}). 
Note that there is generally no simple one-to-one relationship 
between eigenmode and a single planet (see text). (a) $g$ modes
(inclinations zero, viewed from ecliptic north; 
$\theta = \vpi_2 - \vpi_1$). The apsidal
precession is prograde (counterclockwise) in the same 
direction as the orbital motion (light blue and red arrows).
(b) $s$ modes. The nodal precession is retrograde (clockwise)
in the opposite direction as the orbital motion (light 
blue and red arrows).
The orbit normals $\nv_i$ hence describe a 
precessional/nutational motion (here about the $z$-axis).
\label{fig:gsmod}
}
\end{figure}

\begin{figure}[t]
\vspace*{-53ex}\hspace*{-07ex}
\includegraphics[scale=0.9]{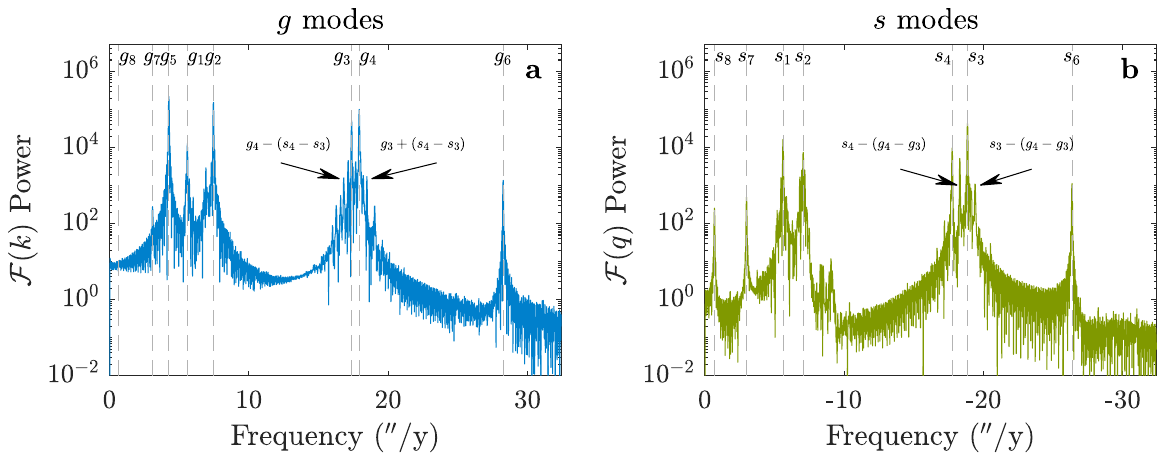}

\vspace*{-50ex}
\caption{\scs
Time series analysis of (a) $k$ and (b) $q$ for Earth 
(see Eqs.~(\ref{eqn:hk}) and~(\ref{eqn:pq}))
to extract solar system secular (or fundamental) frequencies from 
numerical integrations (solution \ZBETa). Time interval: [$-$20 0]~Myr. 
\calF\ = Fast-Fourier Transform (FFT).
Vertical dashed lines indicate frequencies (see Table~\ref{tab:freqs}).
Note that $s$ frequencies (b) are negative (retrograde
nodal precession).
Also note significant side peaks due to \gftL\ around the main 
peaks $s_3$ and $s_4$.
Side peaks around $g_3$ and $g_4$ in (a) are also present
due to \sftL, illustrating the interaction of $g$
and $s$ modes. For a ratio $\sftL:\gftL =\ $2:1, $s_4 - \gftL =
s_3 + \gftL$.
\label{fig:gsfft}
}
\end{figure}

The secular frequencies of the solar system
naturally split into $g$ and $s$ modes,
which are loosely related to the apsidal and nodal precession of 
the orbits, respectively (see Fig.~\ref{fig:gsmod}).
The $g$ and $s$ frequencies are constant in 
quasi-periodic systems but vary over time in chaotic systems
and as such are critical for understanding the long-term
behavior of the solar system.
The $g$'s and $s$'s may be obtained by spectral analysis 
of the classic variables for each planet
(see Fig.~\ref{fig:gsfft} and Table~\ref{tab:freqs}):
\beqn
h =  e \sin(\vpi)         \quad & ; & \quad
k =  e \cos(\vpi)         \label{eqn:hk} \\
p = \sin (I/2) \ \sin \Om \quad & ; & \quad
q = \sin (I/2) \ \cos \Om \label{eqn:pq} \ ;
\eeqn
for $e$, $I$, $\vpi$, and $\Om$, see Figs.~\ref{fig:eccInc},
\ref{fig:gsmod}, 
and Table~\ref{tab:notval}. In the Laplace-Lagrange solution,
the $h, k, p$, and $q$ can be written as simple trigonometric 
sums with constant $g$'s and $s$'s \citep[e.g.,][]{murraydermott99}.

Keeping in mind the generally complex relation between individual
planets and eigenmodes, the $g$ and $s$ modes may be schematically 
illustrated (see Fig.~\ref{fig:gsmod} for $N = 3$). The $g$ modes 
involve variations in $e$ and \vpi, where $e$ usually varies between 
some extreme values (black double arrows, Fig.~\ref{fig:gsmod}a) and
\vpi\ characterizes the apsidal precession; \vpi\ may librate (oscillate) 
or circulate for solar system orbits. In the linear system and for 
a simple eigenmode-planet relationship, the time average
$\langle\vpi_i\rangle$ may be written as $\langle\vpi_i\rangle \simeq
g_i t$, or $\langle\dot{\vpi}_i\rangle \simeq g_i$ (where ``dot'' 
denotes the time derivative).
The planetary $g_i$'s ($i = 1,\ldots,8$)
are positive (see Table~\ref{tab:freqs}), hence for circulating \vpi, 
the time-averaged apsidal precession is prograde
(i.e., in the same 
direction as the orbital motion, see large arrows in 
Fig.~\ref{fig:gsmod}a). The frequency $g_9$ (dominated by
Pluto) is negative (Table~\ref{tab:freqs}). 
Importantly, however, the values and signs of
the solar system's fundamental frequencies are not deduced 
from the motion of individual planets, but from the eigenmode
analysis of the full system.

\renewcommand{\baselinestretch}{\blss}\selectfont
\begin{table}[h]
\caption{Solar system secular (or fundamental) frequencies (arcsec y\pmo\
= \asy)$^a$ and periods (y) from \ZBETa. Interval: [$-$20 0]~Myr. 
\label{tab:freqs}}
\vspace*{5mm}
\hspace*{-5mm}
\begin{tabular}{crrrrr}
\hline
\#  & $g$    & $T_g$   & $s$         & $T_s$    &  \\
& $^b$(\asy) & (y)     & (\asy)      & (y)      &  \\
\hline \\[0ex]
1 &      5.5821 &    232,170 &   $-$5.6144 &    230,837 & \\
2 &      7.4559 &    173,821 &   $-$7.0628 &    183,498 & \\
3 &     17.3695 &     74,613 &  $-$18.8476 &     68,762 & \\
4 &     17.9184 &     72,328 &  $-$17.7492 &     73,017 & \\
5 &      4.2575 &    304,404 &      0.0000 &        $-$ & \\
6 &     28.2452 &     45,884 &  $-$26.3478 &     49,188 & \\
7 &      3.0878 &    419,719 &   $-$2.9926 &    433,072 & \\
8 &      0.6736 &  1,923,992 &   $-$0.6921 &  1,872,457 & \\
9 &   $-$0.3494 &  3,709,721 &   $-$0.3511 &  3,691,356 & \\
\hline
\end{tabular}
\noindent {\scs \\[2ex]
$^a$ 1~arcsec = 1/3600 of a degree. \\
$^b$ Frequency conversion from kyr\pmo\ to arcsec~y\pmo\
     is ($\x 3600 \cdot 360/1000 = 1296$).
} 
\end{table}
\renewcommand{\baselinestretch}{\bls}\selectfont

The lines of apsides of two orbits coincide, i.e., their perihelia
are aligned, for $\theta = \vpi_2 - \vpi_1 = 2\pi \cdot n$, where $n$ 
is an integer (see Fig.~\ref{fig:gsmod}a), or in the simple, linear 
system at $(g_2 - g_1) t = 2\pi \cdot n$, i.e., at a frequency 
($g_2$$-$$g_1$). Thus, one might expect that 
a major component in a planet's orbital eccentricity 
spectrum occurs at the difference between two secular frequencies
(relative motion). Indeed, the highest power in, e.g., Earth's 
recent orbital eccentricity 
spectrum occurs at about $3.2~\asy$ (arcsec~y\pmo,
see Table~\ref{tab:gij}), or a period of \sm{405}~kyr
(see Fig.~\ref{fig:speceH}). 
It turns out that this frequency is associated with \gtfL;
the values given in Table~\ref{tab:freqs} yield
$\gtfL = 7.4559 - 4.2575 = 3.1984$~\asy, or 405~kyr.
The \gtfL\ cycle is also called long eccentricity cycle
(see Section~\ref{sec:einc}) and represents the most stable $g$ mode 
combination term in Earth's eccentricity spectrum in the recent past
\citep[although for deep time, see][]{zeebelantink24aj}. 
The various combinations of 
($g_i$$-$$g_j$) and ($s_i$$-$$s_j$) relevant for Earth's orbital
eccentricity and inclination are discussed in 
Sections~\ref{sec:fcomb} and~\ref{sec:amfm}.

The $s$ modes 
involve variations in $I$ and \Om, where $I$ usually varies between 
some extreme values (black double arrows, Fig.~\ref{fig:gsmod}b) and 
\Om\ characterizes the nodal precession; \Om\ may librate 
or circulate. In the simple, linear system, the time average
$\langle\Om_i\rangle$ may be written as $\langle\Om_i\rangle \simeq
s_i t$, or $\langle\dot{\Om}_i\rangle \simeq s_i$. 
The planetary $s_i$'s are negative (see Table~\ref{tab:freqs}), 
hence for circulating \Om, the time-averaged nodal precession is 
retrograde
(i.e., in the opposite direction as the orbital motion, 
see large arrows in Fig.~\ref{fig:gsmod}b). Given conservation
of total angular momentum ($\vb{L}$), there exists an invariable 
plane perpendicular to $\vb{L}$, which is fixed in space. It
follows that one of the $s$ frequencies is zero ($s_5$, see 
Table~\ref{tab:freqs}). Thus, the $p$ and $q$ solutions behave
slightly different than the $h$ and $k$ solutions (see 
Eqs.~(\ref{eqn:hk}) and~(\ref{eqn:pq})). The eccentricity of an
orbit introduces a reference line and hence an asymmetry into the 
problem for the $g$ modes ($h, k$), which is not the case for 
the $s$ modes ($p, q$). For the latter, the mutual inclination 
between, say, two orbits matters, not the absolute inclination
relative to an arbitrary reference plane \citep{murraydermott99}.

\subsubsection{Frequency Combination Terms: Eccentricity \&
Obliquity-Modulation \label{sec:fcomb}}

\begin{figure}[t]
\vspace*{-40ex}\hspace*{+00ex}
\includegraphics[scale=0.8]{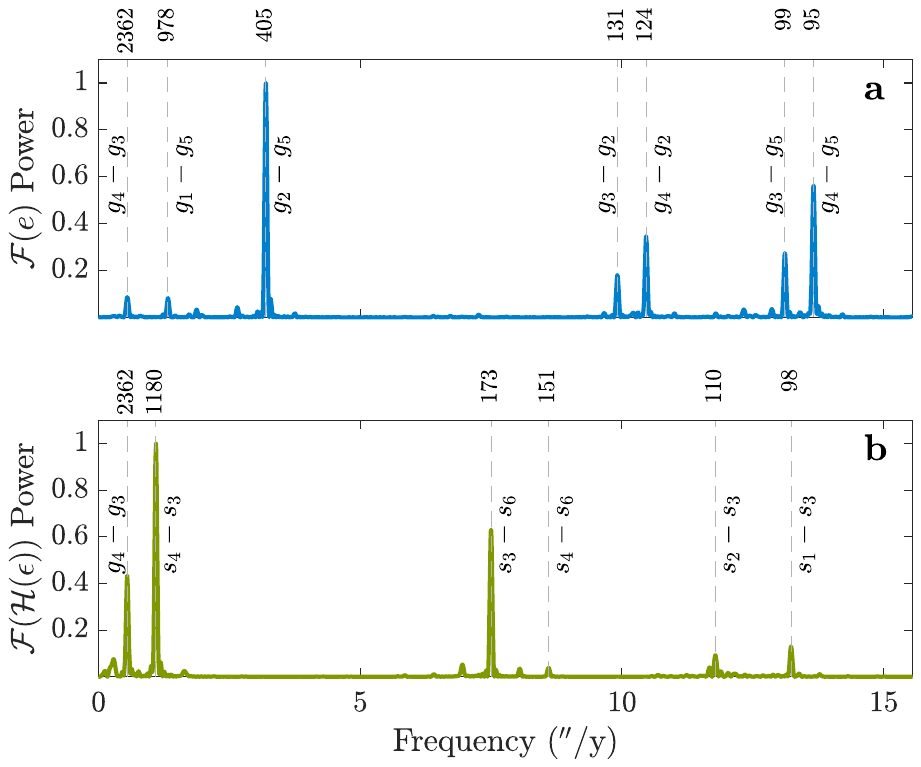}

\vspace*{-30ex}
\caption{\scs
(a) FFT of Earth's orbital eccentricity in the recent
past ($\calF(e)$). Interval: [$-$20 0]~Myr.
(b) FFT of the Hilbert transform of Earth's obliquity
($\calF(\calH(\obl))$.
Spectral power is normalized to maximum power.
Numbers above panels indicate periods in kyr.
\label{fig:speceH}
}
\end{figure}

As discussed above, major components (terms) in relevant orbital 
parameter spectra usually occur at the difference between 
pairs\footnote{Multiple frequency combinations usually
lead to minor terms, some of which may be observable in the 
geologic record \citep[for a recently noticed \sm{200}-kyr 
cycle, see][]{hilgen20}. of secular frequencies
(relative motion).} For instance, the main eccentricity
terms in the recent past (405~kyr and \sm{100}~kyr) are 
due to combination pairs of
$g$ frequencies (see Fig.~\ref{fig:speceH}a and 
Table~\ref{tab:gij}). Importantly, the \sm{100}~kyr band
is actually composed of four individual frequencies,
which, in geologic sequences, are identifiable separately 
however only in high-quality cyclostratigraphic records.
The $g$-frequency combination terms that dominate orbital 
eccentricity (Fig.~\ref{fig:ei}) are directly accessible 
through time series analysis of eccentricity
(here FFT, $\calF(e)$, Fig.~\ref{fig:speceH}). Note
that in general, there is no analogous relationship between 
$s$-frequency combinations and
inclination, again due to the dependence of inclination 
on the reference frame (see Section~\ref{sec:einc}).
A natural choice for extracting $s$-frequency combinations 
from inclination would be the invariant frame. However, orbital
elements in the invariant frame are not universally 
provided by \ASs. Moreover, compared to eccentricity, 
there is also no equivalent expression of inclination 
in stratigraphic sequences that could be used for practical 
purposes.

A more useful approach for extracting $s$-frequency 
combinations from \ASs\ is analyzing the amplitude 
modulation of obliquity. This approach also makes practical 
sense because obliquity and its AM are often 
expressed in cyclostratigraphic records. To obtain the 
AM frequency terms of obliquity from \ASs, we first extract 
the envelope of obliquity by applying the Hilbert transform, 
\calH(\obl) (see also Fig.~\ref{fig:am-g43})
and then apply FFT ($\calF(\calH(\obl))$, 
Fig.~\ref{fig:speceH}b and Table~\ref{tab:gij}).
The strongest obliquity AM terms in the recent past 
are \sftL\ and \stsL, which are discussed in more detail 
in Section~\ref{sec:amfm}.
The next highest peak is due to $g$ modes (not $s$ modes),
i.e., the combination \gftL, which may be surprising if one 
expects only $s$ terms to appear in the obliquity AM
spectrum. However, as shown above, significant \gftL\ 
side peaks are present around $s_3$ and $s_4$
(Fig.~\ref{fig:gsfft}). In turn, the combination of the main-
and side 
peaks then lead to significant power at the beat frequency 
\gftL\ in the obliquity AM spectrum. The interaction of
$g$ and $s$ modes, i.e., $g$ frequencies appearing in 
$q$'s spectrum and $s$ frequencies appearing in $k$'s
spectrum (see Eqs.~(\ref{eqn:hk}),~(\ref{eqn:pq}),
and Fig.~\ref{fig:gsfft}) illustrate the non-linear
behavior of solar-system dynamics (in the linear system,
the $g$ and $s$ modes are decoupled and independent
of each other). In fact, \sftL\ and \gftL\ are currently in 
a so-called 2:1 resonance state (periods of \sm{1.2} and 
\sm{2.4}~Myr, Table~\ref{tab:gij}), which is a major contributor 
to long-term solar-system chaos (see Sections~\ref{sec:chaos} 
and~\ref{sec:amfm}).

\renewcommand{\baselinestretch}{\blss}\selectfont
\begin{table}[h]
\caption{Frequency terms (arcsec y\pmo\ = 
\asy)$^a$ and periods (kyr) based on $g$ and $s$ frequencies in \ZBETa\
(interval: [$-$20 0]~Myr, see Table~\ref{tab:freqs}). 
\label{tab:gij}}
\vspace*{5mm}
\hspace*{-5mm}
\begin{tabular}{ccrrl}
\hline
Term           & Short-    & Freq.      & Period    & Expressed \\
               & hand      & $^b$(\asy) & $^c$(kyr) & in        \\
\hline \\[0ex]
($g_4$$-$$g_3$) & $g_{43}$ &  0.5487 &  2362 & Ecc.\& Oblq. \\
($g_1$$-$$g_5$) & $g_{15}$ &  1.3246 &   978 & Eccentricity \\
($g_2$$-$$g_5$) & $g_{25}$ &  3.1984 &   405 & Eccentricity \\
($g_3$$-$$g_2$) & $g_{32}$ &  9.9137 &   131 & Eccentricity \\
($g_4$$-$$g_2$) & $g_{42}$ & 10.4624 &   124 & Eccentricity \\
($g_3$$-$$g_5$) & $g_{35}$ & 13.1121 &    99 & Eccentricity \\
($g_4$$-$$g_5$) & $g_{45}$ & 13.6609 &    95 & Eccentricity \\
                &          &  ------ &       &              \\
($s_4$$-$$s_3$) & $s_{43}$ &  1.0983 &  1180 & Obliquity-AM \\
($s_3$$-$$s_6$) & $s_{36}$ &  7.5003 &   173 & Obliquity-AM \\
($s_4$$-$$s_6$) & $s_{46}$ &  8.5986 &   151 & Obliquity-AM \\
($s_2$$-$$s_3$) & $s_{23}$ & 11.7847 &   110 & Obliquity-AM \\
($s_1$$-$$s_3$) & $s_{13}$ & 13.2331 &    98 & Obliquity-AM \\
\hline
\end{tabular}
\noindent {\scs \\[2ex]
$^a$ 1~arcsec = 1/3600 of a degree. \\
$^b$ Frequency conversion from kyr\pmo\ to arcsec~y\pmo\
     is ($\x 3600 \cdot 360/1000 = 1296$). \\
$^c$ Uncertainties in \gftL\ and \sftL\ periods
     from spectral analysis may be 
     estimated as $\pm 40$~kyr and $\pm 10$~kyr \citep{zeebe22aj}. \\
} 
\end{table}
\renewcommand{\baselinestretch}{\bls}\selectfont

\subsection{The so-called ``Metronomes'' \label{sec:mtrn}}

The term astrochronological {\it metronome} usually refers to a
prominent and exceptionally stable frequency in Earth's 
orbital parameters that may be widely used to construct accurate 
chronologies.
Importantly, for the frequencies to be of use in practical 
applications requires both frequency and amplitude to be stable.
The eccentricity term \gtfL\ (\sm{405}~kyr in the recent past) and
the inclination term \stsL\ (\sm{173} in the recent past)
have been labeled \rev{the} eccentricity- and inclination 
metronome, respectively \citep{laskar20gts}.
However, \citet{zeebelantink24aj} have recently shown
that \gtfL\ can become unstable on long time scales,
which compromises the 405-kyr cycle's reliability beyond several 
hundred million years in the past. The term \stsL\ is also 
unreliable, yet on even shorter time scales. For example,
analysis of the obliquity modulation (illustrated in 
Fig.~\ref{fig:speceH}) of the solution \ZBETa\ shows that 
the dominant term involving $s_6$ across the 
interval $-$66 to $-$56~Myr
is \sfsL, and not \stsL. Note that \ZBETa\ has been
geologically constrained to $-$66~Myr, which
includes the interval in question from $-$66 to $-$56~Myr
\citep[see][and 
Table~\ref{tab:os}]{zeebelourens22epsl}. Shifts in the dominant
terms are due to changes in $s_3$ and $s_4$, which are known 
to be variable beyond \sm{50}~Myr (see Fig.~6 of 
\citet{zeebelourens22epsl}). As a result, the
dominant period involving
$s_6$ may shift to \sm{151}~kyr or even alternate
over time between \sm{173} and \sm{151}~kyr. In summary,
the current evidence does not support the notion of
generally stable and prominent metronomes for universal 
use in astrochronology and
cyclostratigraphy. \rev{The term \stsL\ is reliable only over 
the past 50~Myr or so, an interval over which the secular 
frequencies appear stable anyway. The term \gtfL\ is more
stable but cannot be taken for granted beyond several 
100~Myr due to solar system chaos.}

\subsection{Solar System Chaos \label{sec:chaos}}

Large-scale dynamical chaos is an inherent characteristic of the 
solar system and fundamental to understanding the behavior of 
astronomical solutions and their limitations, which is described
below. First, an example of a simple \rev{analog} mechanical system 
is introduced to illustrate some basic aspects of chaos.

\subsubsection{The Driven Pendulum: Poincar{\'e} Section
\label{sec:pend}}

Fundamental contributions to the theory of dynamical 
systems (including the gravitational three-body problem
and the solar system) were made by Henri Poincar{\'e} (1854-1912), 
whose work laid the foundations of chaos theory.
Poincar{\'e} wrote:
``It may happen that small differences in the initial conditions 
produce very great ones in the final phenomena. A small error in 
the former will produce an enormous error in the latter. Prediction 
becomes impossible $\ldots$'' \citep{poincare1914}. 
The sensitivity to small differences in initial conditions
or small perturbations is a key element of chaotic systems.

\begin{figure}[t]
\def\sc{0.63}
\vspace*{-40ex}                          
\includegraphics[scale=\sc]{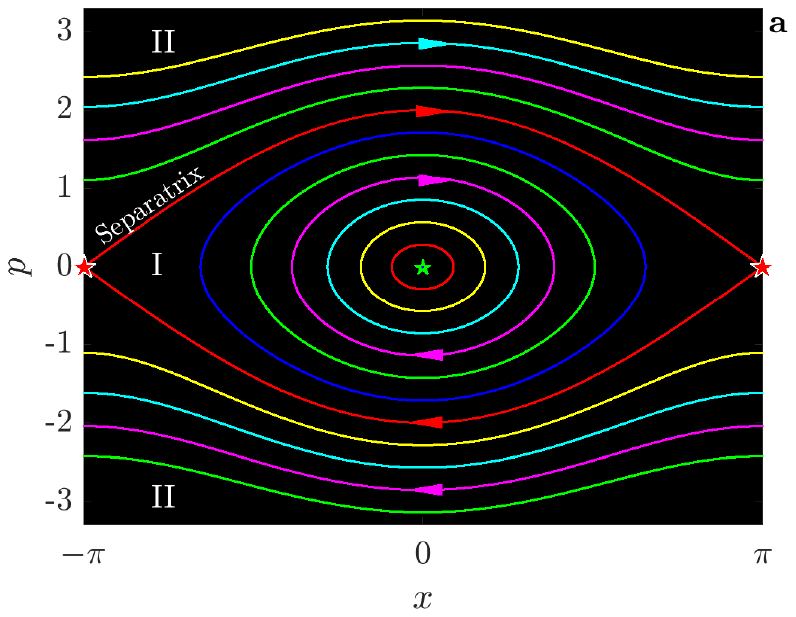}

\vspace*{-58.5ex}
\includegraphics[scale=\sc]{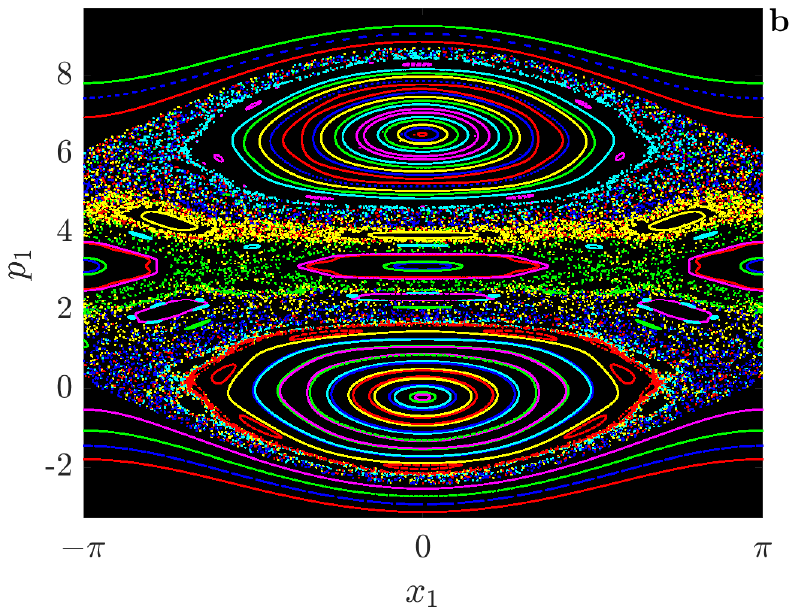}

\vspace*{-59.5ex}
\includegraphics[scale=\sc]{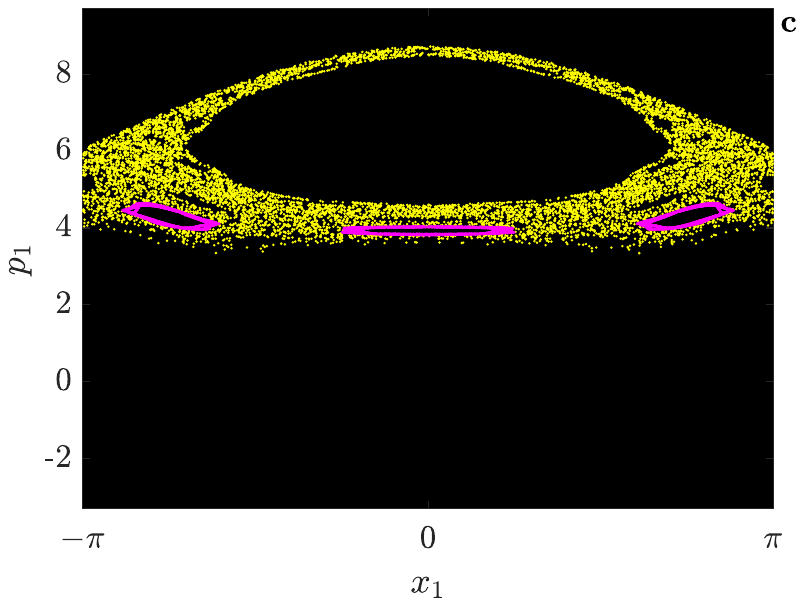}

\vspace*{-30ex}
\caption{\scs
(a) Phase diagram ($x =$ position/angle, $p =$ momentum)
of the \rev{simple} rigid pendulum (units normalized). 
Green/red stars: stable/unstable fixed points (modulo
$2\pi$). Color coding indicates phase space 
trajectories for each set of initial conditions ($x^0,p^0$),
$N = 16$, from which the equations of 
motion are numerically integrated. 
(b) Poincar{\'e} section (aka surface of section) of the
driven \rev{rigid} pendulum ($N = 64$ initial conditions)
\citep[simplified Hamiltonian, see Chapter~6,]
[]{morbidelli02}.
(c) Same as (b) for $N = 2$, $p^0_1$ differs by $10^{-6}$.
\label{fig:chaos}
}
\end{figure}

A simple mechanical system often used to illustrate
such behavior is the driven \rev{rigid} pendulum
\citep[e.g.,][]{chirikov79,lichtenberg83,murray01,morbidelli02}. 
Consider first a simple rigid pendulum in a gravitational 
field with angle $x$ to the vertical
(equivalent to position) and 
momentum $p = m l \dot x$, where $m$ and $l$ are its effective 
mass and length, respectively (for phase diagram, see 
Fig.~\ref{fig:chaos}a). For small to moderate energies 
(regime~I), the pendulum librates around the 
equilibrium point ($x,p$) = (0,0), representing a single
oscillator. The trajectories 
originating in regime~I are considered to be in resonance
\citep{murray01}.
For large energies (regime~II), 
the pendulum rotates clockwise ($p < 0$) or counterclockwise 
($p > 0$). The structure that separates the two regimes is 
called the separatrix (Fig.~\ref{fig:chaos}a). For initial 
conditions starting on the separatrix, the pendulum 
approaches the unstable fixed point (Fig.~\ref{fig:chaos}a,
red star) as $t \rightarrow 
\infty$ (upright position). The motion
of the pendulum starting in the vicinity of initial conditions 
($x^0,p^0$) in regime I or II far from the separatrix is 
quite simple. Starting at neighboring points in regime~I
will lead to libration, whereas starting at neighboring points 
in regime~II will lead to rotation. However, for
($x^0,p^0$) very close to the separatrix, neighboring starting
points may lead to fundamentally different outcomes governing
their motion, e.g., libration {\sl or} rotation. 
In chaotic systems, one may think of the separatrix
as ``the cradle of chaos'' \citep{murray01}.

While the motion of the simple rigid pendulum is non-chaotic
(universally), this is not the case for the driven \rev{rigid} pendulum 
(Fig.~\ref{fig:chaos}b). Imagine a driven pendulum by,
for instance, applying a periodic force (torque) to the 
pendulum or periodically moving its pivot point, which
results in the interaction of two oscillators
\rev{(pendulum and pivot)}.
Instead of two phase-space variables for the \rev{simple} rigid 
pendulum, four variables
are now needed to describe the motion, say $x_1,p_1,x_2$
and $p_2$. Thus, the phase space is four-dimensional (4D), 
which cannot be fully represented in 2D. To
visualize the dynamics, a lower-dimensional
phase diagram is often constructed,
called a Poincar{\'e} section, or surface of section. 
For example, for the driven pendulum, one may select
the surface ($x_1,p_1$), with $x_2$ perpendicular
to the surface, and plot ($x_1,p_1$) only when $x_2 = 0$, 
i.e., at times when the phase trajectory 
intersects the ($x_1,p_1$) plane (see Fig.~\ref{fig:chaos}b).
\rev{Usually, an additional condition is specified for $p_2$, 
e.g., $p_2 > 0$.}
One can think of a Poincar{\'e} section as a stroboscopic 
image of the trajectory's time evolution in phase space 
\citep{morbidelli02}.

The critical ingredient for chaotic dynamics are overlapping 
resonances. For illustration, consider the single
resonance (Fig.~\ref{fig:chaos}a, one `cat's eye') 
and the two resonances (Fig.~\ref{fig:chaos}b, two
`cat's eyes' atop one another).
For certain parameter values
of the driven pendulum (e.g., driving frequency close to the
natural pendulum frequency), the two resonances are close
enough to overlap (Fig.~\ref{fig:chaos}b, around $p_1 = \pi$). 
The structures of certain phase space regions still 
appear relatively simple, such as in the vicinity of the
resonances centered at (0,0) and (0,$\sm{2}\pi$)
and for large $|p_1|$, corresponding to regimes~I and~II,
respectively (Fig.~\ref{fig:chaos}a).
The motion in these regions may be restricted to invariant
(stable) tori, or KAM tori\footnote{
KAM refers to the Kolmogorov-Arnold-Moser theorem, 
demonstrating the persistence of quasi-periodic motion
on invariant tori (singular:~torus)
under small perturbations of, e.g., 
Hamiltonian systems \citep{kolmogorov54,arnold63,
moser62}.
} \citep[for summary, see][]{morbidelli02}.
However, for certain initial conditions, chaotic phase 
space regions appear, expressed in the Poincar{\'e} 
section as disordered sets of scattered points (see, e.g., 
transition areas between regions of libration and rotation,
Fig.~\ref{fig:chaos}b and~c).
In chaotic regions of the phase space, nearby trajectories 
diverge exponentially, characterized by their Lyapunov
exponent (see Section~\ref{sec:intro}). 
As a result, the system is sensitive to small 
differences in initial conditions and small
perturbations. For example, two trajectories of
the driven pendulum ($x^0_1,p^0_1) = (0,\sm{3.82})$
that initially differ by only
$10^{-6}$ in $p^0_1$ rapidly diverge and occupy
different (separate) regions of the phase space
(Fig.~\ref{fig:chaos}c).
Prediction becomes impossible, as Poincar{\'e} had 
pointed out.

The driven pendulum serves as a useful illustration
for various properties of chaotic systems, including
the sensitivity to initial conditions.
The full orbital dynamics of, for instance, 
the solar system, are of course substantially more complex
(see below). Nevertheless, key elements such as
the sensitivity to initial conditions and
the overlap of resonances are also critical in
solar system dynamics (note that for 
secular resonances, the $g$ and $s$ frequencies are
involved, see Section~\ref{sec:gs}). For example,
it was recently 
discovered that Earth's long eccentricity cycle can be 
become unstable on long time scales due to the resonance
$(g_1 - g_2) + (s_1 - s_2) \simeq 0$
\citep[see][]{zeebelantink24aj}

\subsubsection{Solar System: Time Scales and Limitations}

As mentioned above, large-scale dynamical chaos is an 
inherent property of the solar system,
which has been independently confirmed in various
numerical studies \citep[e.g.,][]{sussman88,
laskar89,ito02,morbidelli02,varadi03,batygin08,zeebe15apje,
brownrein20,hernandez22,abbot23,zeebelantink24aj}.
\rev{For observational studies, see, e.g.,
\citet{mameyers17,westerhold17,olsen19,zeebelourens19}.}
Dynamical chaos 
affects the secular frequencies $g_i$ and $s_i$ (see 
Section~\ref{sec:gs}), where the terms \gftL\ and \sftL, for instance,
show chaotic behavior already on a 50-Myr time scale. As a 
result, astronomical solutions diverge around $t = \pm50$~Myr,
which fundamentally prevents identifying a unique solution on 
time scales $\gsim$10$^8$~y \citep{laskar04Natb,zeebe17aj,
zeebelourens19}. 
The quest for a single deterministic solution, which conclusively 
describes the solar system's evolution for all times \citep[in the 
spirit of Laplace's demon, see][]{laplace51}, must therefore be 
regarded as quixotic. In fact, long-term predictability is
fundamentally unachievable \citep{poincare1914}.
The chaos not only severely limits
our understanding and ability to reconstruct and predict the 
solar system's history and long-term future, it also imposes 
fundamental limits on geological and astrochronological 
applications such as developing a fully calibrated astronomical 
time scale beyond \sm{50}~Ma, \rev{including the SEC}
\citep[for recent efforts, see][]
{zeebelourens19,zeebelourens22epsl,kocken24}. 

\begin{figure}[t]
\vspace*{-33ex}\hspace*{-02ex}
\includegraphics[scale=0.80]{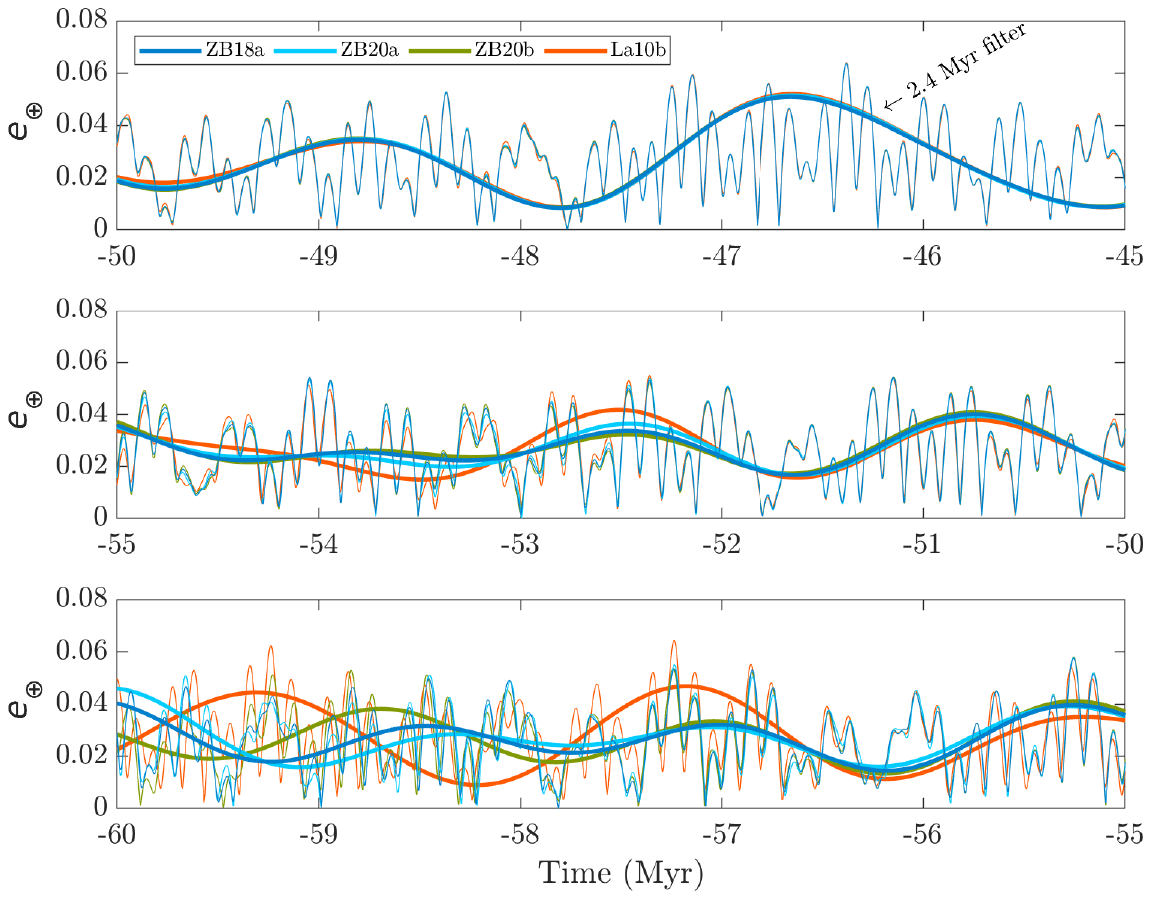}

\vspace*{-25ex}
\caption{\scs
Earth's orbital eccentricity (\eE) from different OS:
\ZBETa\ \citep{zeebelourens19},
\ZBXXa\ and \ZBXXb\ \citep{zeebelourens22epsl},
and \LaXb\ \citep{laskar11}.
The 2.4-Myr filter is sensitive to the amplitude modulation 
of eccentricity, with minima corresponding to a reduced
amplitude of the short eccentricity cycle, aka very long 
eccentricity nodes (VLN).
\label{fig:soln}
}
\end{figure}

The limits imposed by dynamical chaos on astronomical
calculations may be illustrated by comparing Earth's
eccentricity (\eE) from different OS prior to $-50$~Myr
(Fig.~\ref{fig:soln}). As an example, we select three of
the most current and up-to-date OS (\ZBETa, \ZBXXa, and \ZBXXb),
which have been constrained by geologic data up to an age of
66~Ma (see Section~\ref{sec:u2dOS}
and \citet{zeebelourens19,zeebelourens22epsl}). 
\ZBETa, \ZBXXa, and \ZBXXb\
use identical initial conditions but feature slightly different
$J_2$ values \citep[for details, see Section~\ref{sec:gr} and][]
{zeebe17aj,zeebelourens19,zeebelourens22epsl} 
and number of asteroids included in 
the simulation. Visually, \eE\ calculated based on these 
three OS is nearly identical back to ca.\ $-58$~Myr but 
diverge quickly beyond that time, as highlighted
by a 2.4-Myr filter (Fig.~\ref{fig:soln}). The 2.4-Myr 
filter is sensitive to the amplitude modulation of
eccentricity, aka very long eccentricity nodes (VLN), with
frequency \gftL. Of the \LaXx\ solutions, \LaXb\ and 
\LaXc\ appear to provide the best match with geological
data to \sm{58}~Ma, but not beyond \citep{westerhold17,
zeebelourens19,zeebelourens22epsl}. \LaXb, for example, 
exhibits a large 100-kyr amplitude just prior
to $-59$~Myr (see Fig.~\ref{fig:soln}), while \ZBETa\ 
and \ZBXXa\ exhibit small 100-kyr amplitudes (VLNs). 
Features such as VLNs are important criteria to
distinguish between different OSs based on geologic
data \citep{westerhold17,zeebelourens19,
zeebelourens22epsl}.

The example using \eE\ based on different OSs 
(Fig.~\ref{fig:soln}) again illustrates one key feature of
dynamical chaos, i.e., the sensitivity to tiny 
perturbations or small differences in initial conditions 
(see Section~\ref{sec:pend}). 
For example, the masses of the largest asteroids are
roughly {\sl ten billion times smaller} than the solar mass
(which represents the magnitude of the first-order 
gravitational interaction). Yet, changing the
number of asteroids included in the simulation
drives the solutions apart beyond ca.\ $-58$~Myr
(Fig.~\ref{fig:soln}).
Small differences in trajectories grow exponentially, 
with a time constant (Lyapunov time, see Section~\ref{sec:intro})
for the inner planets of $\sim$3-5~Myr estimated
numerically \citep{laskar89,varadi03,batygin08,zeebe15apj}.
For example, a difference in initial position
of $1$~cm grows to $\sim$1~AU (=~1.496\e{11}~m) after 
90-150~Myr, which makes it fundamentally impossible to 
predict the evolution of planetary orbits accurately 
beyond a certain time horizon.

\begin{figure}[t]
\vspace*{-37ex}\hspace*{+08ex}
\includegraphics[scale=0.65]{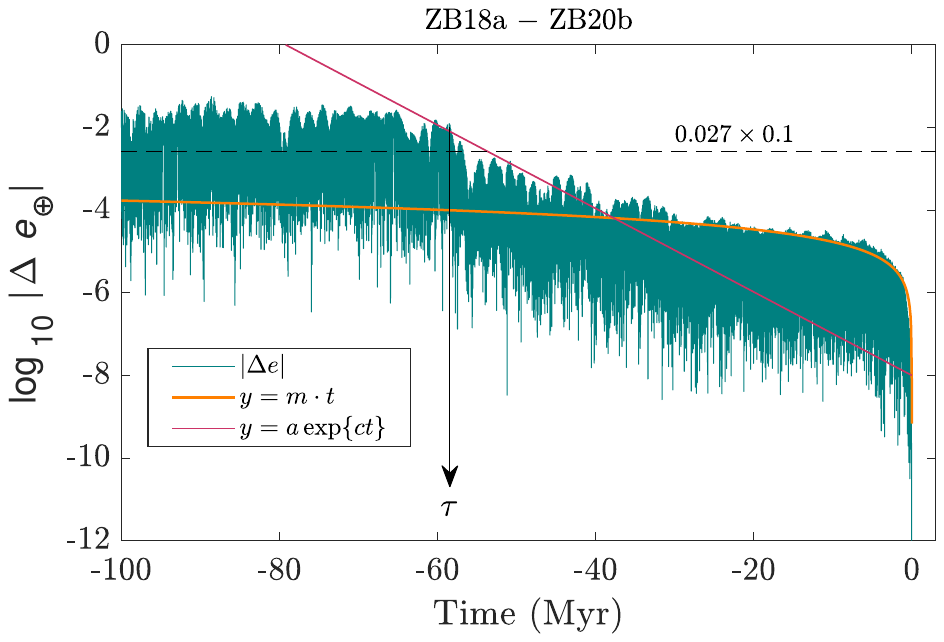}

\vspace*{-30ex}
\caption{\scs
Difference in Earth's orbital eccentricity on a log-$y$ scale, 
$\log_{10}|\DeE|$,
between the solutions \ZBETa\ and \ZBXXb\
over the past 100~Myr. The arrow indicates
the divergence time $\tau$, when max$|\DeE|$ irreversibly 
crosses \sm10\% of mean \eE\ ($\sm0.027\x0.1$, dashed horizontal
line). Orange curve: simple fit function with linear growth
in $|\DeE|$. Red line: exponential growth in $|\DeE|$ 
(linear on log-$y$ scale) with a Lyapunov time of
4.3~Myr (see text).
\label{fig:tauIll}
}
\end{figure}

In contrast to the limitations discussed above
(largely due to unstable terms 
such as \gftL\ and \sftL), another frequency term appears
more promising as it shows more 
stable behavior. For example, it has hitherto been assumed that
\gtfL\ was practically stable in the past and has been suggested
for use as a ``metronome'' in deep-time geological applications, 
i.e., far exceeding 50~Ma \citep{laskar04Natb,kent18,
spalding18,meyers18,montenari18,lantink19,devleesch24}.
The \gtfL\ cycle, which is the dominant term in Earth's orbital
eccentricity in the recent past (\sm{405}~kyr, see 
Fig.~\ref{fig:speceH}) may thus have been regarded as an 
island of stability in a sea of chaos. However, 
as mentioned above, \gtfL\ can als become unstable over 
long time scales \citep{zeebelantink24aj}.

\subsubsection{Divergence Time $\tau$}

The action of chaos on Earth's orbital evolution may be 
illustrated by the difference between two eccentricity
solutions ($|\DeE|$), say plotted on a log scale (Fig.~\ref{fig:tauIll}).
Starting at $t_0$ and integrating backwards in time,
integration errors dominate initially. Integration errors
usually grow polynomially (for angle-like variables), 
here $\propto \ t$ and dominant 
for $t \ \gsim -40$~Myr (see orange line, Fig.~\ref{fig:tauIll};
note that a function
linear in $t$ appears non-linear on a log scale).
Ultimately, the divergence of trajectories is dominated by 
exponential growth ($t \ \lsim -40$~Myr), which is indicative 
of chaotic behavior (red line, Fig.~\ref{fig:tauIll}).
For $t \ \lsim -60$~Myr, the solutions have completely
diverged and their difference may reach the maximum
of \sm{0.06} ($\max|\DeE|$ no longer increases,
see Fig.~\ref{fig:soln}).
The difference between two orbital solutions may be 
tracked by the divergence time $\tau$, i.e., the time interval 
after which the absolute 
difference in Earth's eccentricity ($|\DeE|$) irreversibly crosses 
\sm10\% of mean \eE\ ($\sm0.027\x0.1$, Fig.~\ref{fig:tauIll}).
Simply put, $\tau$ represents the time interval beyond which the 
solutions no longer agree. The divergence time is largely 
controlled by the Lyapunov time, although the two are different 
quantities.

\subsubsection{\rev{Chaos and Inapplicability of Basic Statistics}}

\begin{figure}[t]
\vspace*{-28ex}\hspace*{+10ex}
\includegraphics[scale=0.55]{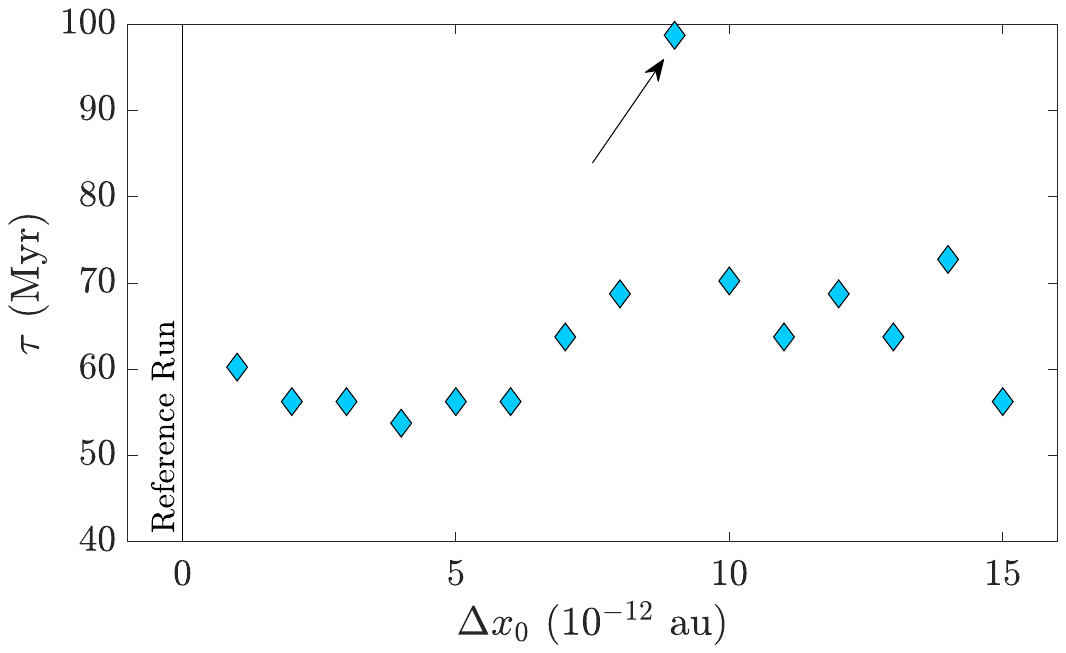}

\vspace*{-25ex}
\caption{\scs
Ensemble integrations with \orb\
\citep[general relativity, $J_2$, and the lunar contribution
turned off, see][]{zeebe23aj}.
Earth's initial $x$-coordinate is perturbed by a small
$\D x_0 = 1\e{-12}$~au (\sm{15~cm}),
relative to the corresponding reference run ($\D x_0 = 0$~au).
The difference 
between two orbital solutions $i$ and $j$ may be tracked using 
the divergence time $\tauD$ \citep[see Fig.~\ref{fig:tauIll} 
and][]{zeebe17aj}.
The run resulting in the largest $\tauD \simeq 100$~Myr 
(arrow) is not a mistake, numerical error, or ``outlier''
(see text).
\label{fig:tauD}
}
\end{figure}

Chaos introduces elements of inherent unpredictability 
into dynamical systems, which generally leads to fundamental
unaccountability in terms of basic statistics.
\rev{Note that the following is not a criticism of statistics,
but rather of attempts to apply basic statistical
concepts to chaotic patterns, which may arise from a
misunderstanding of the dynamics of the system.}
For instance, consider a dynamical system parameter, which, 
when increased by a small amount \del\ produces a change in a 
system variable by an amount $X$. If \del\ is now 
varied incrementally, one may expect
that $X$ is a function of \del\ (linear or 
nonlinear) that somehow scales with the size of \del.
However, this is generally not the case in chaotic systems
(see Section~\ref{sec:pend}).
As an example, consider the solar system in which Earth's 
initial position ($x$-coordinate) is varied by 
$\del = \D x_0$. The system is then integrated for
different $\D x_0$ over 100~Myr and the differences in
Earth's orbital eccentricity \eE\ are evaluated
\citep[][]{zeebe23aj}, say tracked using the divergence 
time \tauD\ (see Fig.~\ref{fig:tauD}). Note 
that \tauD\ is not a measure of accuracy or inaccuracy.

First, there is no systematic relationship between
\tauD\ and $\D x_0$ that somehow scales with the 
size of $\D x_0$. Second, the run resulting in the
largest $\tauD \simeq 100$~Myr (see arrow) is not a
mistake, numerical
error, or ``outlier'' that can therefore be excluded 
(the run was carefully examined). The run is a proper 
solution of the system, illustrating the unpredictability
and unaccountability of chaotic systems in 
terms of basic statistics.
\rev{For example, values that fall outside two or three 
sigma from the mean are often considered outliers, which is
not applicable here.}
One important corollary is that it is generally not 
possible to ``tune'' or ``fit'' astronomical solutions to 
geological data. The best that can probably be done
is to ``constrain'' astronomical solutions by geological 
data. In other words, to create large ensembles of solutions
via parameter variations and select/discard those 
solutions that show good/poor agreement 
with the data \citep{zeebelourens19,zeebelourens22epsl}.

As outlined in the preceding sections, solar system chaos 
is key to understanding the limitations imposed on orbital 
solutions by chaotic dynamics, resonances, the behavior of 
ensemble integrations, and more.
Chaos also causes critical changes in the amplitude modulation
of orbital forcing signals, which, if expressed in 
cyclostratigraphic sequences, can be used to reconstruct the 
solar system's chaotic history. Elements of amplitude (and 
frequency) modulation are covered in Section~\ref{sec:amfm}.

\subsection{Amplitude \& Frequency Modulation (AM \& FM)
\label{sec:amfm}}

\begin{figure}[t]
\vspace*{-55ex} \hspace*{-12ex}
\includegraphics[width=1.2\linewidth]{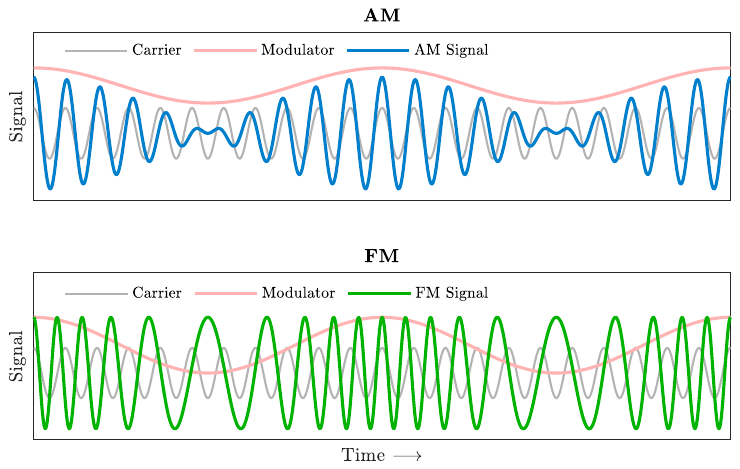}

\vspace*{-52ex}
\caption{\scs
Schematic illustration of amplitude \& frequency 
modulation (AM \& FM).
\label{fig:amfm}
}
\end{figure}

Amplitude and frequency modulation of orbital forcing
signals is similar to AM/FM techniques used in radio 
broadcasting. Schematically, the carrier signal 
(high frequency) is modulated, resulting either in
slow modulation of the amplitude (AM) or frequency
(FM), see Fig.~\ref{fig:amfm}. 
For example, the astronomical carrier frequency
may be precession, eccentricity, or obliquity, while
the AM frequency (or ``beat'', see below) may be eccentricity or 
various combinations of ($g_i$$-$$g_j$) and ($s_i$$-$$s_j$), see 
Section~\ref{sec:fcomb}.
As described below, frequency modulation is less commonly 
used than amplitude modulation in astrochronology and
cyclostratigraphy. Also, while there is a direct analogy 
between radio AM and astronomical forcing AM, the analogy 
is not straightforward for the FM case study discussed here 
(Section~\ref{sec:fm}).

\subsubsection{AM \label{sec:am}}

\begin{figure}[t]
\vspace*{-60ex} \hspace*{-10ex}
\includegraphics[width=1.2\linewidth]{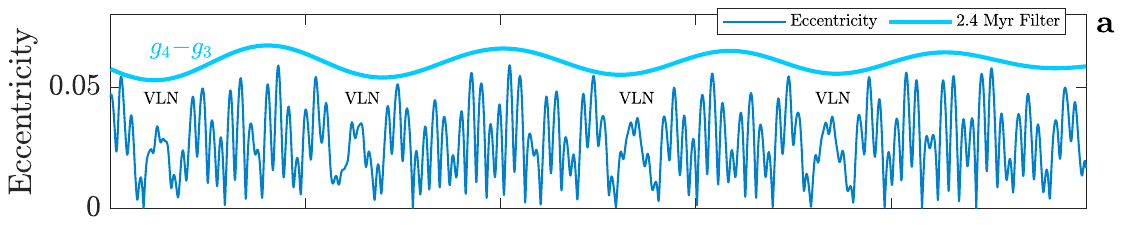}

\vspace*{-120ex} \hspace*{-10ex}
\includegraphics[width=1.2\linewidth]{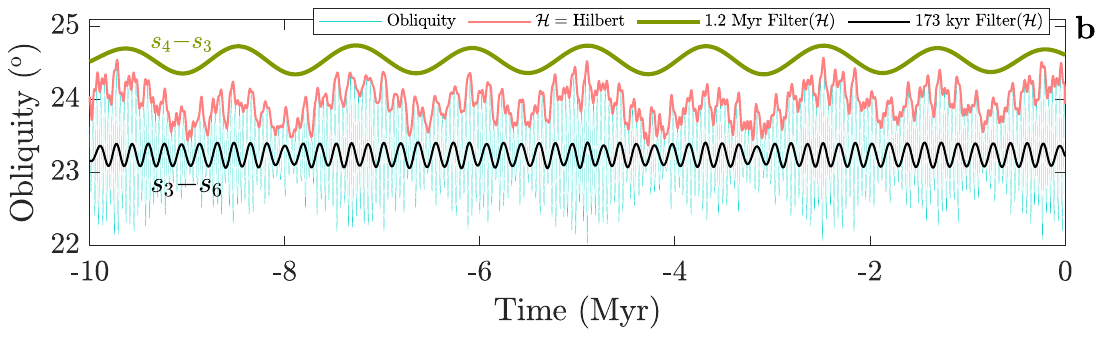}

\vspace*{-58ex}
\caption{\scs
(a) Earth's orbital eccentricity and \sm{2.4}~Myr filter
corresponding to \gftL. 
VLN = very long eccentricity node.
Filters are plotted using arbitrary amplitudes and
vertical offsets.
(b) Obliquity, its Hilbert transform \calH(\obl), \sm{1.2}~Myr 
filter of \calH(\obl)\ corresponding to \sftL, and 173~kyr 
filter of \calH(\obl)\ corresponding to \stsL.
\label{fig:am-g43}
}
\end{figure}

Amplitude modulation is frequently used in cyclostratigraphy
because AM cycles are often expressed in geologic sequences
(also called ``beats'' in analogy to beat frequencies
in music). For example, the amplitude of
climatic precession is modulated by eccentricity
(in a very simple manner, i.e.,
eccentricity is the envelope of climatic precession,
$\pbar = e \sin \ombar$, see Fig.~\ref{fig:eop}).
Also, the SEC's amplitude is modulated by the LEC.
Bundles of four SECs in the recent past (each \sm{100}~kyr,
for details see Eq.~(\ref{eqn:gft})) 
represent one LEC (405~kyr, see Fig.~\ref{fig:e123}a). 
On even longer time scales, the eccentricity amplitude
is modulated by \gftL, which leads to very long 
eccentricity nodes (VLNs) with much reduced SEC amplitude
and a recent period of \sm{2.4}~Myr (Fig.~\ref{fig:am-g43}a).
The spectrum of Earth's orbital eccentricity is largely 
dominated by simple combinations of the $g$ frequencies $g_2$
to $g_5$ (see Tables~\ref{tab:freqs} and~\ref{tab:gij}).
Thus, there is a simple relationship between, for instance,
\gftL\ and the other relevant $(g_i-g_j)$ terms
(individual SEC's are \sm{124}, 131, 95, and 99~kyr):
\beqn
\gftL\pmo\ 
& = & (1/72.328 - 1/74.613)\pmo = 2362~\mbox{kyr} \nn \\
& = &
(g_{42} - g_{32})\pmo = (g_{45} - g_{35})\pmo     \nn \\
& \simeq &
(1/123.87 - 1/130.73)\pmo \simeq (1/94.87 - 1/98.84)\pmo \ .
\label{eqn:gft}
\eeqn
Importantly, owing
to chaotic dynamical behavior, the fundamental frequencies
(and hence the beats) are not constant, but change
over time. The changing VLN patterns beyond about $50$~Ma
observed in cyclostratigraphic
records are important criteria to distinguish between different 
orbital solution \citep{westerhold17,zeebelourens19,
zeebelourens22epsl}.

As discussed above, there is no direct analogy between 
eccentricity and inclination (see Section~\ref{sec:einc}).
Similarly, the analogy between their amplitude modulations (i.e.,
between ($g_i$$-$$g_j$) terms and ($s_i$$-$$s_j$) terms) is limited
as well. The orbital parameter closely related to inclination 
is obliquity, which is often expressed in cyclostratigraphic
sequences. Thus, to examine the AM associated with ($s_i$$-$$s_j$) 
terms, it is helpful to examine the AM of obliquity (see 
Fig.~\ref{fig:am-g43}b). Again, to do so, we first extract the envelope
of obliquity by applying the Hilbert transform, \calH(\obl).
The two largest peaks in \calH(\obl)'s 
spectrum are located at periods of \sm{1.2}~Myr
and \sm{173}~kyr (see Fig.~\ref{fig:speceH}). 
The 1.2-Myr period corresponds to \sftL,
which is currently in a 2:1 resonance with \gftL\
(Section~\ref{sec:fcomb}). On time scales beyond \sm{50}~Myr, the 
\sftL:\gftL\ ratio is subject to chaotic transitions and
may exhibit a wide range of values on Gyr-time scale
\citep{zeebelantink24pa}.
The 173-kyr period corresponds to \stsL, which is 
significant in the recent past
\citep[e.g.,][]{shack99,hinnov00,paelike04},
and has been suggested to be expressed in multiple proxy 
records of Mesozoic and Cenozoic age \citep[for recent summary 
see, e.g.,][]{huang21}. However, for records older than 
\sm{50}~Ma, the assumed assignment and period is ambiguous
and hence problematic
owing to the lack of generally stable and prominent 
metronomes or geochronometers, including \stsL\ 
(see Section~\ref{sec:mtrn}).

\subsubsection{FM \label{sec:fm}}

\begin{figure}[t]
\vspace*{-40ex} \hspace*{-00ex}
\includegraphics[width=1.0\linewidth]{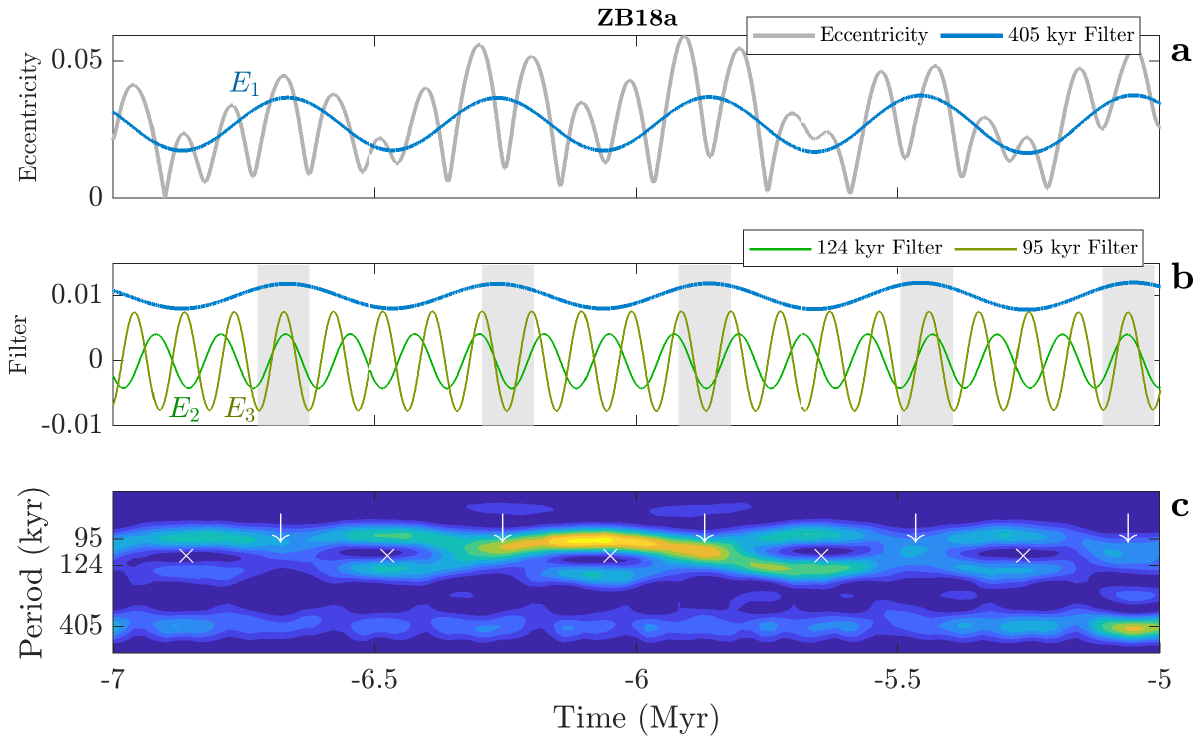}

\vspace*{-32ex}
\caption{\scs
Illustration of eccentricity-related FM interference patterns.
(a) Earth's orbital eccentricity (\eE) from solution \ZBETa\ and 
405-kyr filter ($E_1$). 
(b) Narrow filters of \eE\ around~124 ($E_2$) and 95~kyr ($E_3$). 
Gray bars mark intervals of $E_2$-$E_3$ constructive (in-phase) 
interference.
Blue: scaled and offset 405-kyr filter from (a).
(c) Evolutive Harmonic Analysis (EHA) of \eE\ (showing 
amplitude, not spectral power) using \ascr\ 
\citep{meyers14ascr}. Arrows and crosses: nodes and rings
in $E_2$-$E_3$ amplitude corresponding to maxima and minima 
in $E_1$, respectively (see (a)).
\label{fig:e123}
}
\end{figure}

Frequency modulation is less commonly used than 
amplitude modulation in cyclostratigraphic studies 
\citep[for a few examples,
see][]{herbert92,liu95,hinnovpark98,rial99,
hinnov00,zeeden15,laurin16,piedrahita22}.
\rev{
One major challenge for cyclostratigraphic applications is 
the conversion from stratigraphic depth to time. Variations in 
sedimentation rate that are not taken into account would perturb 
the frequency of recorded astronomical cycles (similar to
the FM illustration shown in Fig.~\ref{fig:amfm}). Conversely, 
tuning of the stratigraphic record to an astronomical frequency 
would map any frequency modulation in the data to variations in 
sedimentation rate. These issues represent substantial obstacles 
to FM analyses of sedimentary data.}
Another potentially complicating factor regarding FM is 
that age-model tuning approaches may introduce systematic 
frequency modulations \citep[e.g.,][]{zeeden15}.
Nevertheless,
analysis of short eccentricity-related FM interference patterns in
stratigraphic sequences can, for instance, be used to discern 
LEC minima and maxima from SEC records \citep[e.g.,][]
{laurin16,piedrahita22}.
The link between the LEC and SECs is of course due to 
the relationship between the $g$ frequencies and
their involvement in Earth's orbital eccentricity, \eE\
(see Sections~\ref{sec:gs} and~\ref{sec:fcomb}). 
The strongest terms in \eE's
spectrum in the recent past are the LEC \gtfL\ (\sm{405}~kyr)
and the SECs \gfwL\ and \gffL\ (\sm{124} and 95~kyr,
$E_2$ and $E_3$ for short) (see Figs.~\ref{fig:speceH}
and~\ref{fig:e123}). 
The combination of the two SECs
in turn yields the LEC ($E_1$):
\beqn
(1/95 - 1/124)\pmo \simeq 405~\mbox{kyr} \ .
\eeqn

Narrow filters of \eE\ around~124 and 95~kyr show
interference patterns that correlate with minima and maxima
in $E_1$. For instance, $E_2$ and $E_3$ interfere 
constructively (in-phase) during $E_1$ maxima and 
destructively (out-of-phase) during $E_1$ minima 
(Fig.~\ref{fig:e123}a and~b). Note that LEC maxima
and minima correlate with large and small SEC amplitudes,
respectively.
Evolutive Harmonic Analysis (EHA) of \eE\ reveals a 
chain pattern in the $E_2$-$E_3$ amplitude, where nodes
(single amplitude and period) correspond to $E_1$ maxima and 
rings (split amplitudes and periods) to $E_1$ minima (see 
arrows and crosses, Fig.~\ref{fig:e123}c).
Thus, the SEC-FM interference patterns can be used 
to deduce LEC minima and maxima \citep{laurin16}.

\subsection{Up-to-Date Orbital Solutions \label{sec:u2dOS}}

As described above, there is a long history of OSs ---
yet only a few are up-to-date and should be employed
in current and future applications, which has not
been the case in many recent applications, indicating
confusion about proper use of OSs (see remark at the
end of this section). 
For the time span from $-300$ 
to 0~Myr, we recommend the solutions \ZBETa\ and \ZBXXx,
which have been constrained by geologic data up to an age of
66~Ma (see Table~\ref{tab:os}).
Importantly, significantly beyond the geologically constrained 
intervals,
or the solution's validity range, the OSs are unconstrained 
owing to solar system chaos. For much older time intervals, the
OSs may serve as examples for possible dynamical patterns
\citep{zeebelantink24aj,zeebelantink24pa}, 
or as potential tuning targets for the LEC in the 
recent past (\sm{405}~kyr). However, the probability that a particular
OS represents the actual, unique history of the solar system
is near zero significantly beyond the OS's validity range. 
For example, at 
double-precision floating-point arithmetic, one could generate 
$10^{21}$ possible different solutions, all within observational 
uncertainties just for Earth's position \citep[ignoring all
other physical and numerical uncertainties in solar system 
models, initial conditions, etc., see][]{zeebe17aj}. 
Regarding geological constraints on OSs based on data-OS
comparisons \citep{zeebelourens19}, note that a bad match 
in a well-constrained younger interval disqualifies a solution 
for {\sl all} older intervals \citep{zeebelourens22epsl}.

\def\s{\scs}
\renewcommand{\baselinestretch}{\blss}\selectfont
\begin{table*}[t]
\caption{Selected (recent or frequently used) orbital solutions.
\label{tab:os}}
\vspace*{5mm}
\hspace*{-5mm}
\begin{tabular}{llllrl}
\hline \\[0ex]
Label  &    & Time Span  & Time Span & Reference    & Notes \\
       &    & Provided   & Valid     &              &       \\
       &    & (Myr)      & (Myr)$^a$ &              &       \\ [1ex]      
\hline \\[-1ex]
\ZBXXtx&    & $-$3500 to 0   & $-$48 to 0 & 
                      \s\citet{zeebelantink24pa}                    & $^b$ \\
\ZBXXx &    & $-$300 to 0    & $-$66 to 0 & 
                      \s\citet{zeebelourens22epsl}                  & $^c$ \\
\ZBETa &    & $-$300 to 0    & $-$66 to 0 & 
                      \s\citet{zeebelourens19,zeebelourens22epsl}   & $^c$ \\
\ZBSTx &    & $-$100 to 0    & $-$50 to 0  & \s\citet{zeebe17aj}    & \\
\LaXx  &    & $-$250 to 0    & $-$50 to 0  & \s\citet{laskar11}     & \\
\Laiv  &    & $-$250 to 250  & $-$40 to 0  & \s\citet{laskar04Natb} & $^d$ \\
\hline
\end{tabular}
\noindent {\scs \\[2ex]
$^a$ Approximate valid time span. \\
$^b$ Ensemble solutions designed for deep-time applications.
     For $-300$ to 0 Myr, \ZBETa\ and \ZBXXx\ are recommended. \\
$^c$ Valid time span indicates geologically constrained time span. \\
$^d$ Outdated compared to \LaXx, \ZBSTx, \ZBETa, and \ZBXXx\ (see text).
} 
\end{table*}
\renewcommand{\baselinestretch}{\bls}\selectfont

The solutions \LaXx, \ZBSTx, \ZBETa, and \ZBXXx\ agree closely
to about $-50$~Myr. However, the \Laiv\ solution does not and is 
only valid to about $-40$~Myr \citep{zeebe17aj,laskar20gts}. 
The physical model 
underlying the \Laiv\ solution did not include asteroids
and used initial conditions from the DE406 ephemeris
\citep[an older ephemeris version created in 1997, see][]{standish98}. 
Given the above and that 
\Laiv\ disagrees beyond $-40$~Myr with the updated solutions 
\LaXx\ from the same group, \Laiv\ is considered outdated. 
Thus, it is clear at this point that \Laiv\ does not
represent a proper solution of the solar system beyond its 
valid time span and should no longer be considered for
any time prior to ca.\ $-40$~Myr. The situation is different 
(i.e., the jury is still out) for e.g., \ZBETa\ and \ZBXXx,
which have not shown inconsistencies with updated/more
recent solutions 
or geologic data \citep{zeebelourens22epsl}.
In other words, \ZBETa\ and \ZBXXx\ could theoretically be 
viable solutions beyond their valid time span, whereas 
\Laiv\ can not.
Unfortunately, there is substantial confusion in the
literature regarding the valid time span of astronomical 
solutions, most notably \Laiv\ \citep[of the numerous examples
only a few recent ones are cited here, e.g.,][]{liu19,
husinec23,wu23,charbonnier23,dutkiewicz24,vervoort24}.

\subsection{Additional Physical/Dynamical Effects in 
Up-to-Date Orbital Solutions \label{sec:gr}}

Generating adequate, state-of-the-art orbital solutions requires
accurate and fast integration of the fundamental dynamical 
equations for the main solar system bodies (see 
Section~\ref{sec:intro}). Furthermore, 
several additional effects need to be considered, some 
(or all) of which have been included in fully numerical solutions 
\citep[e.g.,][]{quinn91,
varadi03,laskar04Natb,laskar11,zeebe17aj,zeebe23aj}.
Additional effects include:
(1) post-Newtonian corrections from general relativity 
\rev{(1PN = first-order post-Newtonian)},
(2) the effect of the Moon,
(3) the Sun's quadrupole moment $J_2$, and 
(4) a contribution from asteroids.
1PN effects are critical \citep[non-negligible,][]
{einstein16} and a fast (symplectic) implementation
is highly desirable for long-term integrations
\citep[see discussion in][]{saha94,zeebe23aj}.
The Moon may be included as a separate object \citep{varadi03,
rauch02,laskar04Natb,zeebe17aj}, or the Earth-Moon system 
may be modeled as a gravitational quadrupole \citep{quinn91,
varadi03,rauch02,zeebe17aj,zeebe23aj}.
The solar quadrupole moment $J_2$ is due to the rotation of 
the Sun, which causes solar oblateness and slightly distorts 
the gravitational field and hence the planetary orbits
\citep[for summary, see, e.g.,][]{rozelot11}. Despite their
comparatively small mass, asteroids are dynamically 
relevant and have been included in short-term, as well as
long-term integrations
\citep[e.g.,][]{standish95,laskar11,zeebe17aj,antonana22}.
Note that while the effects from $J_2$ and asteroids
may appear negligible, their contributions become critical
for astronomical solutions over, e.g., 50-Myr time scale due 
to solar system chaos (Section~\ref{sec:chaos}).
The effect of solar mass loss over 50~Myr may be
neglected \citep{zeebelourens19} but needs to be considered
in deep time \citep[e.g.,][]{minton07,spalding18,zeebelantink24aj}.

%
\section{Precession-Tilt Solutions \label{sec:pt}}

\begin{figure}[t]
\vspace*{-10ex} \hspace*{+10ex}
\includegraphics[scale=0.85]{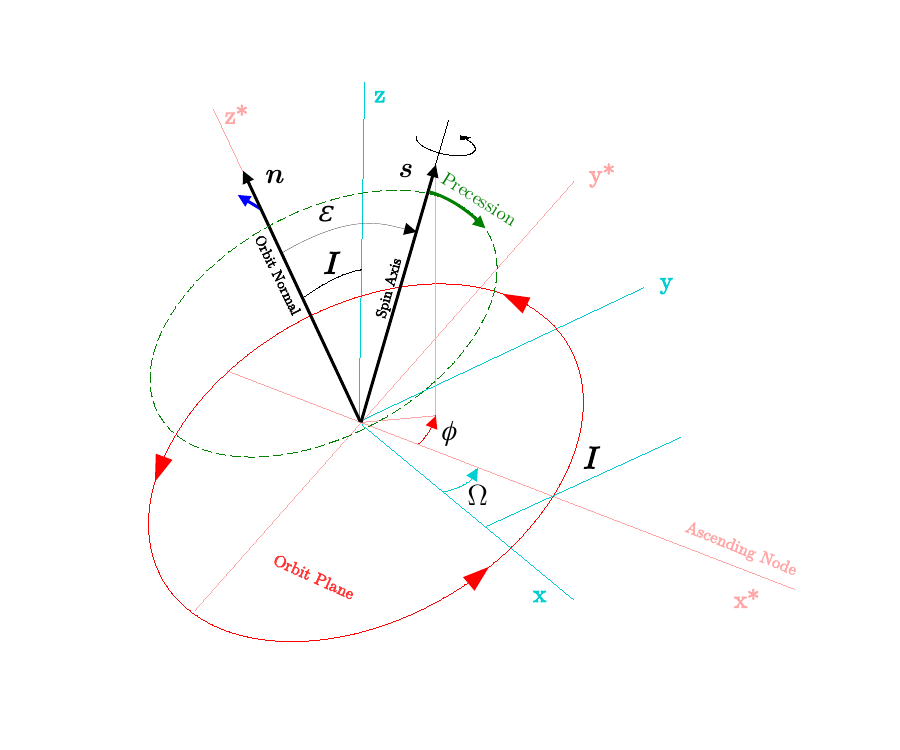}

\vspace*{-05ex}
\caption{\scs
Inertial (fixed) coordinate system ($x,y,z$, light blue) and coordinate 
system moving with the orbit plane ($x^*,y^*,z^*$, red),
with $z^*$ parallel to the orbit normal \nv\ and $x^*$ along 
the ascending node. \sv\ is the spin axis, the obliquity
(axial tilt)
\obl\ is the angle between \sv\ and \nv\ ($\cos\obl = \nv 
\cdot \sv$). \Om\ is the longitude of the ascending node,
$I$ is the orbital inclination (angle between $z^*$ and $z$),
and $\prc$ measures the precession angle in the orbit plane
(see text).
\sv\ slowly precesses westward (clockwise) along the ecliptic 
in the opposite direction (retrograde, green arrow and green
dashed line) to Earth's orbital motion (red arrows) and spin.
The planetary (or ecliptic) precession is indicated by the 
small blue arrow (see text).
\label{fig:illprc}
}
\end{figure}

The second type of astronomical solutions discussed here 
(in addition to orbital solutions), are Precession-Tilt 
(PT) solutions. As mentioned above, precession and tilt
(= obliquity) are among the most frequently used orbital parameters 
in the Earth sciences (see Fig.~\ref{fig:eop}). Defining 
obliquity (\obl) is straightforward, that is, \obl\ 
is the angle between Earth's spin axis and the orbit normal, i.e., 
unit vectors \vb{s} and \vb{n}, where \vb{n} is  
perpendicular to Earth's orbital plane 
= ecliptic (see Fig.~\ref{fig:illprc}). Equivalently, obliquity
is the angle 
between the celestial equator and the ecliptic. In contrast, 
the term {\it precession} may be used interchangeably in the 
literature for different concepts/motions. Moreover, the geometric 
description, as well as the different notations and symbols used 
by different authors, is often confusing. Importantly, however,
an accurate description of precession is inevitably somewhat
more complex because of the various motions and angles
involved. Most critical for cyclostratigraphic and 
astrochronological applications is being aware of the 
differences between climatic and luni-solar precession,
as well as nodal and apsidal precession 
(Section~\ref{sec:gs}).

\subsection{Luni-Solar and Climatic Precession 
\label{sec:lscp}}

\begin{figure}[t]
\vspace*{-35ex} \hspace*{+00ex}
\includegraphics[scale=0.7]{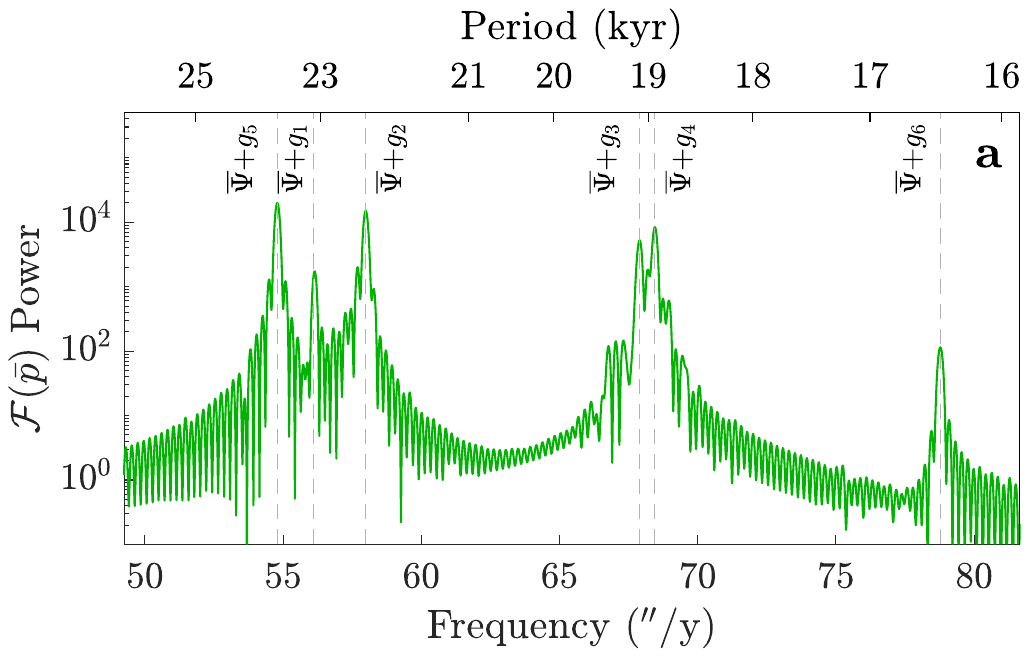}

\vspace*{-62ex} \hspace*{+00ex}
\includegraphics[scale=0.7]{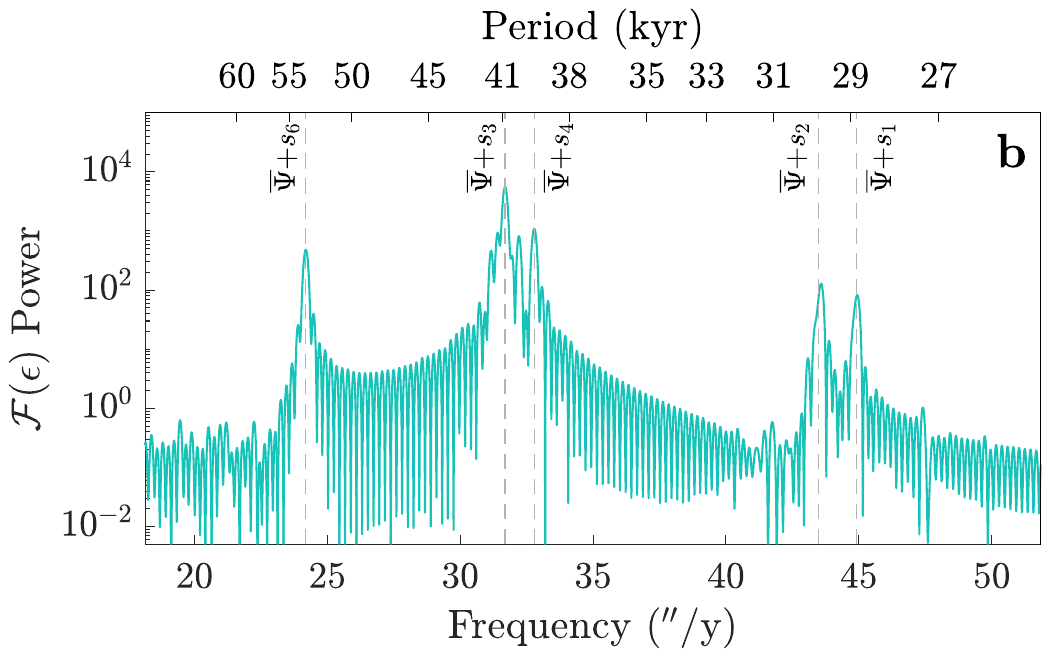}

\vspace*{-30ex}
\caption{\scs
(a) FFT$\ = \calF(\pbar)$ of climatic precession ($\pbar = e \sin 
\ombar$) and (b) FFT$\ = \calF(\obl)$ of obliquity, both
from \snv\ output based on \ZBETa(1,1) (see 
Section~\ref{sec:EdTd}) over a 6-Myr interval ($-6$ to 0~Myr).
\label{fig:cpfft}
}
\end{figure}

At present, the Earth rotates on its axis
at an angular velocity of 
\sm{7.29\e{-5}}~rad~s\pmo\ (see spin axis vector \vb{s},
Fig.~\ref{fig:illprc}). The spin axis itself slowly precesses
westward (clockwise, precession angle \prc, Fig.~\ref{fig:illprc}) 
along the ecliptic in the opposite direction (retrograde, 
green arrow) to Earth's orbital motion 
and spin (red arrows, see also Fig.~\ref{fig:eop}).
The axial precession (rate) along the fixed ecliptic with respect 
to space is referred to as {\it luni-solar precession}
\citep{williams94}. The present precession rate of the total 
space motion at $t_0$ is $\Psin \simeq 50.38$~\asy\ \citep[][]
{capitaine03}, corresponding to a period of \sm{25.7}~kyr
($\Psi$ represents a frequency, i.e., rate of change
= time derivative, see Section~\ref{sec:eqec};
conventionally, $\Psi$ is taken positive).
Over time, the axial precession therefore causes
a notable motion of the equinoxes westward along the ecliptic.
When it comes to astronomical forcing of Earth's climate, 
however, it is not the position of the equinoxes 
relative to the fixed stars that matters ($\prc$ alone)
but the position relative to Earth's elliptic orbit
(see Section~\ref{sec:milank}),
i.e., the perihelion \citep[for further illustrations, 
see, e.g.,][]{hinnov18,lourens21}. In addition, the
forcing depends on orbital eccentricity (in a circular orbit where 
$e = 0$, for example, the equinox position is inconsequential).
Climate forcing is thus controlled by the {\it climatic precession}
\pbar\ (see Fig.~\ref{fig:eop}):
\beqn
\pbar = e \sin \ombar \ ,
\eeqn
where $\ombar = \vpi + |\prc|$ is the longitude of 
perihelion measured from the moving equinox (see below for \prc's 
sign; note also that various symbols are used for \ombar\ in 
the literature). 
As a result, precessional climate forcing does not
follow a simple sine function with 25.7-kyr periodicity but 
depends on the frequency spectra of \vpi\ and \prc, with
its amplitude modulated by $e$ (see Figs.~\ref{fig:eop}
and~\ref{fig:cpfft}).

Given that the precession rate is related to \PSI\
and \vpi's rate is related to the g-mode frequencies $g_i$ 
(see Section~\ref{sec:gs}), one might expect that a spectral 
analysis of the climatic precession would show peaks at 
$\Psibar + g_i$, where \Psibar\ is the mean \PSI\ across
the analyzed interval. For retrograde axial precession
and prograde apsidal precession, the frequencies add up
to a higher frequency. Indeed \pbar's power spectrum
shows dominant peaks, for instance, at \sm{19}, \sm{22.4}, and
\sm{23.7}~kyr, corresponding to $\Psibar + g_i$ where 
$i = 3,4,2$ and~5 (Fig.~\ref{fig:cpfft}). Analogous to
$\Psibar + g_i$, which yields the precession frequencies,
$\Psibar + s_i$ yields the obliquity frequencies.
Note that the dominant $g_i$ are positive and the dominant 
$s_i$ are negative. For retrograde axial precession and retrograde 
nodal precession, the frequencies combine to a lower frequency.
Thus, precession and obliquity frequencies are higher and lower 
than the luni-solar precession rate, respectively 
(inversely for the periods, see 
Table~\ref{tab:fpt} and Figs.~\ref{fig:cpfft} 
and~\ref{fig:AllT}).

\renewcommand{\baselinestretch}{\blss}
\begin{table}[t]
\caption{Frequencies (ordered by amplitude)
of climatic precession (\pbar) and obliquity (\obl).$^a$ 
\label{tab:fpt}}
\begin{tabular}{llll}
\hline
Frequency       & Value & Period &  Note \\
                & (\asy)& (kyr)  &       \\
\hline
$\Psibar + g_5$ & 54.79 & 23.65 & \\
$\Psibar + g_2$ & 57.99 & 22.35 & \\
$\Psibar + g_4$ & 68.45 & 18.93 & \\
$\Psibar + g_3$ & 67.90 & 19.09 & \\
$\Psibar + g_1$ & 56.14 & 23.08 & \\
$\Psibar + g_6$ & 78.78 & 16.45 & \\
                & ------ &      & \\
$\Psibar + s_3         $ & 31.69 & 40.90 &      \\
$\Psibar + s_4         $ & 32.79 & 39.52 &      \\
$\Psibar + s_3 + g_{43}$ & 32.21 & 40.23 & $^b$ \\
$\Psibar + s_6         $ & 24.19 & 53.57 &      \\
$\Psibar + s_3 - g_{43}$ & 31.17 & 41.58 & $^b$ \\
$\Psibar + s_1         $ & 44.94 & 28.84 &      \\
\hline
\end{tabular}

{\scs
$^a$ From FFT of \snv\ output based on \ZBETa(1,1) (see 
Section~\ref{sec:EdTd}) over 6-Myr interval ($-6$ to 0~Myr).
Note that identifying $\Psibar + s_2$ in \obl's
spectrum is obscured due to several large $s_2$ side peaks
(see Fig.~\ref{fig:gsfft}) that are unresolved over a 6-Myr 
interval. \\
$^b$ $g_{43} = \gftL$. Note $g$-$s$ interaction \citep[see][]{zeebe22aj}.
}
\end{table}
\renewcommand{\baselinestretch}{\bls}

Different approaches may be used to compute PT solutions. 
Some methods are based on series expansions or integration
of equations for the 
angles involved \citep[e.g.,][]{sharaf67,
berger76ast,kinoshita75,kinoshita77,berger89ast,laskar93}. 
Other methods
are based on integrating equations for Earth's spin vector 
\citep[e.g.,][]{goldreich66,quinn91,touma94,zeebe22aj}. The latter
approach allows physical insight into the torques involved
(see~\ref{sec:kbet}), 
provides a simple error metric for accuracy during spin vector
integration, and can be applied using the most up-to-date 
orbital solutions (see Sections~\ref{sec:u2dOS} 
and~\ref{sec:u2dPT}). Numerical routines to compute PT
solutions based on Earth's spin vector are freely available 
in C (\snv) and R (\snvR) at \snvurl\ and \snvurlR.

As noted above,
axial precession is retrograde, that is, the equinoxes move 
in the opposite
direction to Earth's orbital motion and spin (see 
Fig.~\ref{fig:illprc}). Given the rule that angles measured 
in the orbital plane are increasing eastward (prograde or
anticlockwise, see Fig.~\ref{fig:eccInc}), it follows
$\dot{\prc}_0 < 0$ \citep{zeebelourens22pa,zeebe22aj}.
Note that the {\it general precession} (usually denoted 
as $p_A$, see below) is often used in this context with 
$\dot{p_A} > 0$ at $t_0$
\citep[e.g.,][]{lieske77,williams94}. For appropriate
initial conditions that refer to the same reference
epoch (e.g., ${p_A}(0) = \prc(0)  = 0$), it follows
$p_A = -\prc$ (see Section~\ref{sec:eqec}).

\subsection{Luni-Solar (Equatorial) and Planetary (Ecliptic)
Precession \label{sec:eqec}}

As mentioned above \citep[and pointed out before, e.g.,][]
{williams94,hilton06},
confusion may also arise over precession
because of different notations and symbols. For example, {\it 
luni-solar precession} and {\it planetary precession} (see below)
 may formally be referred 
to as {\it precession of the equator} and {\it precession of 
the ecliptic}, respectively \citep{fukushima03,hilton06}. The 
two quantities are critical for understanding the link between 
solar system dynamics and spin axis dynamics, and hence between 
OS and PT solutions.

The luni-solar precession rate (here symbol \PSI, see 
Table~\ref{tab:notval}) at $t_0$ ($\Psin > 0$)
may be calculated 
from \citep[e.g.,][] {ward74,quinn91,williams94}:
\beqn
\Psin = \alp \cos \obl_0 + \ggp \ ,
\label{eqn:psi}
\eeqn
where \alp\ \rev{is often referred to as precession 
``constant'' (\ref{sec:kbet}, although 
cf.~Section~\ref{sec:psilong})},
$\obl_0$ is the 
obliquity angle at $t_0$, and \ggp\ is the geodetic precession 
\citep[for details, see][]{quinn91,zeebe22aj}. 
Eq.~(\ref{eqn:psi}) strictly only applies at $t_0$
and does not capture certain periodic variations
(with long-term zero averages though). Note 
that \PSI\ is a rate
(unit \asy), not an angle (the angle is usually denoted 
as $\Psi_A$, index `$A$' for accumulated). 
\PSI\ refers to the total precessional
space motion in an inertial coordinate system 
\citep[fixed ecliptic, see][]{williams94}. However, for many 
applications, the motion relative
to the moving ecliptic may be relevant (not to be confused with
the moving equinoxes, see Section~\ref{sec:lscp}). As 
discussed in Section~\ref{sec:os}, the planetary orbits are not
stationary but move in space over time due to mutual 
interactions. Thus, the motion of Earth's orbital plane 
(ecliptic) contributes to the precessional 
motion (planetary or ecliptic precession, see 
Fig.~\ref{fig:illprc}, small blue arrow). The precession relative 
to the moving ecliptic is referred to as {\it general 
precession in longitude}
($\pA = -\prc$, see Fig.~\ref{fig:illprc}), or just 
{\it general precession}. Note that \pA\ is an angle; 
its rate is $d \pA /dt = \dot{p}_A$. In 
summary, with respect to \prc, we may write:
\beqn
\phi = -p_A \qq ; \qq \dot{\prc} = - \dot{p}_A \ .
\eeqn
Briefly, the important difference between
\prc\ and $\dot{\prc}$ vs.\ \PSI\ and $\dot{\Psi}$
is that the first two are measured in the moving 
ecliptic frame, while the last two are measured
in the inertial frame.

\begin{figure}[t]
\vspace*{-45ex} \hspace*{+00ex}
\includegraphics[scale=0.7]{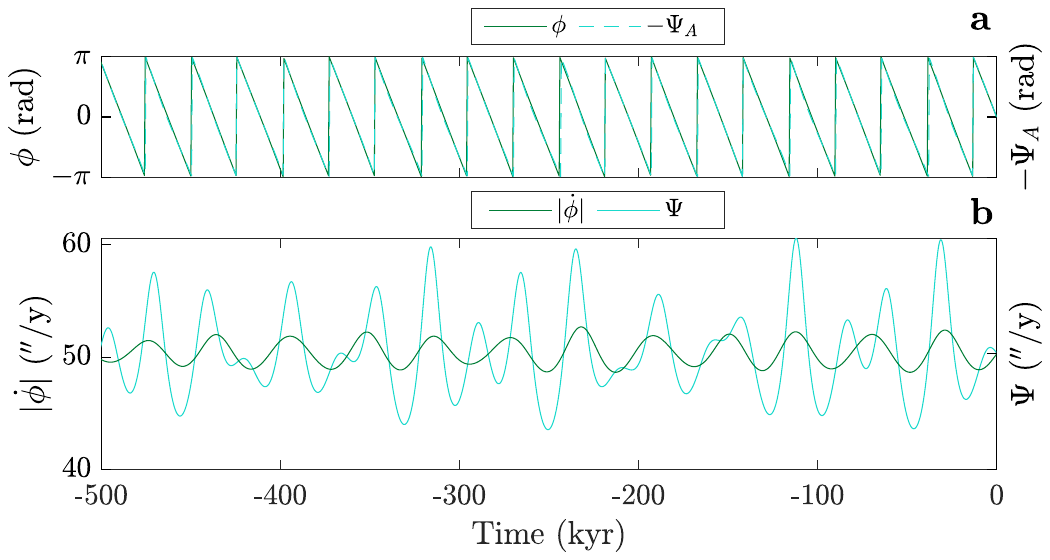}

\vspace*{-30ex}
\caption{\scs
(a) Precession angle \prc\ (in moving ecliptic frame,
see Fig.~\ref{fig:illprc}) and $-\PSI_A$ (in inertial frame); 
both angles are taken as zero at $t_0$.
(b) Time derivative of precession angle $|\dot{\prc}|$ 
(in moving ecliptic frame) and luni-solar precession
\PSI\ (in inertial frame).
$\Psi$ is conventionally taken positive.
Note that \PSI\ is not constant over time and the difference
$|\dot{\prc}| - \PSI$ (ecliptic motion) is small at $t_0$,
but not in general.
\label{fig:phi}
}
\end{figure}

Once $\dot{p}_A$ and \PSI\ and have been calculated,
the precession components
are often separated into a luni-solar and planetary 
contribution. At $t_0$, the difference
$\dot{p}_A - \PSI$ ({\it ecliptic 
motion}) may be written as \citep{williams94}: \\[-3ex]
\beqn
\dot{p}_{A,0} - \Psin   & = &
50.2880       - 50.3848   =   - 0.0968
          \qq \mbox{[\asy]} \ ,
\label{eqn:ecm}
\eeqn
where \citet{capitaine03}'s values have been used 
(see Table~\ref{tab:notval}). Similar values have been
obtained based on a variety of approaches 
\citep[e.g.,][]{lieske77,laskar93,williams94,simon94,roosbeek98,
capitaine03,fukushima03,hilton06,vondrak11}.
From Eq.~(\ref{eqn:ecm}) it may appear that the ecliptic
motion is much smaller than the general motion. However,
Eq.~(\ref{eqn:ecm}) only applies to the ecliptic vs.\ 
general motion at $t_0$ and not generally at all times
(see Fig.~\ref{fig:phi} and \citet{vondrak11}).

For certain paleo-applications such as deep-time studies, 
the details and differences between the various precessional 
quantities discussed above may not matter because the long-term 
means of, e.g., \PSI, $|\dot{\prc}|$, and Eq.~(\ref{eqn:psi}) 
are very similar. However, it is clear that \PSI, for instance,
is not constant over time but varies periodically on Milankovi\'c 
time scales and exhibits long-term secular trends 
(Figs.~\ref{fig:phi} and~\ref{fig:AllT}). 
Moreover, its precise value depends on the precessional
model employed and the dynamical ellipticity value at $t_0$ 
(see Section~\ref{sec:EdTd}). Thus, providing precessional 
parameters or \Psin\ with a large number of digits appears 
unnecessary for practical applications (except for check values 
in numerical routines).

\rev{
\subsubsection{\rev{Luni-Solar Precession: Long-term Variations}
\label{sec:psilong}}
Considering only short-term variations in \PSI\ (see 
Eq.~(\ref{eqn:psi})),
the parameter \alp\ may be indeed be taken as constant
(see~\ref{sec:kbet}). However, considering long-term variations 
in \PSI, \alp\ is far from constant due to the long-term evolution 
of the Earth-Moon system, including changes in Earth's 
rotation, lunar distance,
tidal dissipation, dynamical ellipticity, etc.\ (see 
Section~\ref{sec:deep}). For instance, ignoring the small
term \ggp\ and setting $\kap = g_L = 1$ (see~\ref{sec:kbet}), 
Eq.~(\ref{eqn:psi}) may be re-written as:
\beqn
\Psi  & = & {\cal C} \ \q{H}{\OmE}
            \left(
            1 + \q{a^3}{a_L^3} \q{m_L}{M_S} 
            \right) \ \cos \obl \ ,
\eeqn
where ${\cal C} = \q{3}{2} \q{GM}{a^3}$, $H$ is the dynamical 
ellipticity (see Section~\ref{sec:EdTd}), \OmE\ is Earth's 
angular speed, $a$ the semi-major axis of its orbit, $a_L$ 
the Earth-Moon distance parameter, and $m_L/M_S$ the lunar 
to solar mass ratio. Furthermore, $H \propto \ \Om_E^2$
(see Section~\ref{sec:EdTd}) and hence:
\beqn
\Psi  & = & {\cal D} \ \OmE
            \left(
            1 + \q{a^3}{a_L^3} \q{m_L}{M_S} 
            \right) \ \cos \obl \ ,
\label{eqn:klong}
\eeqn
where ${\cal D}$ may be taken as constant (to first order)
over long time scales. Eq.~(\ref{eqn:klong}) illustrates 
the large secular decrease in \PSI\ over geologic time
(i.e, increase in period,
see Fig.~\ref{fig:AllT}) due to the slow-down of Earth's rotation 
($\OmE$) and the simultaneous increase in lunar distance ($a_L$).
The two processes are coupled via conservation of angular 
momentum \citep[e.g.,][]{goldreich66,touma94,farhat22,
zeebelantink24pa,malinverno24}.
Finally, note that several recent studies have estimated 
long-term variations in luni-solar precession frequency 
from stratigraphic data \citep[see, e.g.,][and references
therein]{wumalinverno24}.
}

\subsection{Dynamical Ellipticity and Tidal Dissipation
\label{sec:EdTd}}

\begin{figure}[t]
\vspace*{-45ex} \hspace*{-12ex}
\includegraphics[width=1.2\linewidth]{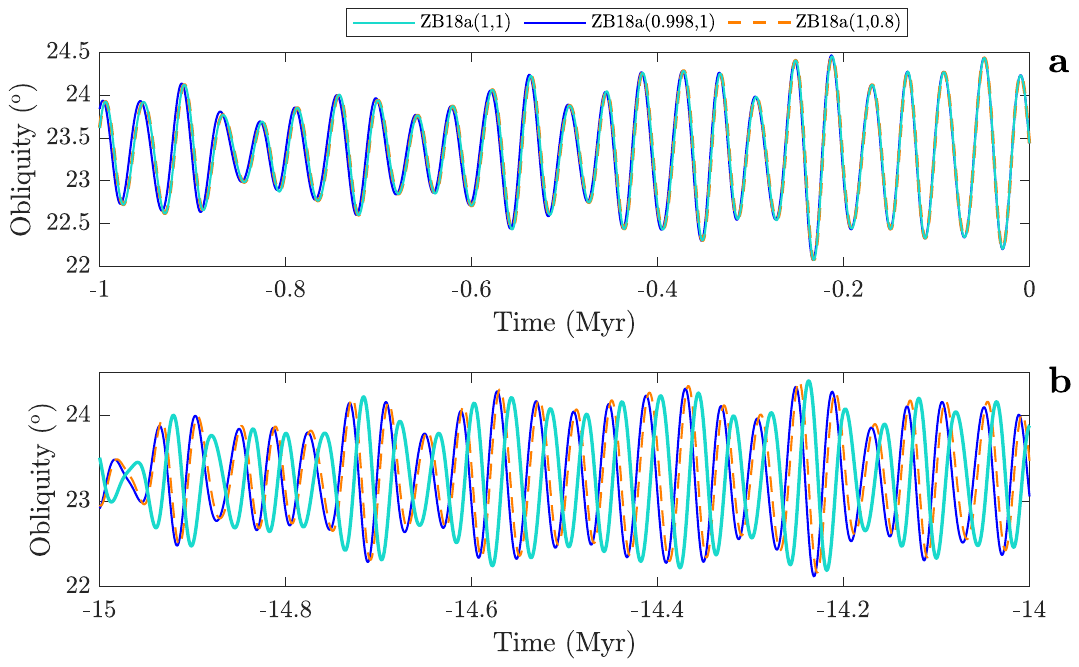}

\vspace*{-40ex}
\caption{\scs
Illustration of changes in Earth's dynamical ellipticity (\Ed) and
tidal dissipation (\Td) affecting the numerical solution
for Earth's obliquity over 15~Myr 
\rev{\citep[for details on the numerical parameters \Ed\ and \Td, see]
[]{zeebelourens22pa}}.
In runs labeled $\Ed = 0.998$ and $\Td = 0.8$, the
values of $\Ed$ and $\Td$ were set to 99.8\% and 80\%, respectively,
of their modern values $\Ed = \Td = 1.0$.
(a) $-1$ to 0~Myr, the solutions are 
nearly identical. (b) $-14$ to $-15$~Myr, the solutions are out of phase,
which would cause significant dating errors.
\label{fig:oblqET}
}
\end{figure}

Dynamical ellipticity refers to the gravitational shape of the Earth, 
largely controlled by the hydrostatic response to its rotation 
rate. Dynamical ellipticity is defined as $H = [C - (A+B)/2] / C$,
where $C$ is the polar moment of inertia and $A$ and $B$ are the 
equatorial moments of inertia. If $A = B$,
\beqn
H = (C - A) / C \qq ; \qq 
H \propto \ \Om_E^2 \ .
\eeqn
Importantly, $H$'s hydrostatic response
is proportional to $\Om_E^2$, where \OmE\ is Earth's 
spin, or angular velocity \citep{lambeck80}.
The present value of $H$ is \sm{0.00328}; 
however, its precise, absolute value is usually adjusted to other 
parameters and hence may differ between models 
\citep[e.g.,][]{quinn91,laskar93,williams94,capitaine03,chen15,
zeebe22aj}.
Note that the calculated changes in obliquity and 
precession in the past are sensitive to even small changes in
dynamical ellipticity (see Fig.~\ref{fig:oblqET}), 
which is affected by, e.g., \OmE, 
ice volume, and mantle convection \citep[e.g.,][]{berger89ast,
laskar93,morrow12,zeebelourens22pa}.
Regarding the effect of Earth's dynamical ellipticity and
Earth's rotation rate
on the luni-solar precession rate (see Eq.~(\ref{eqn:psi})), 
it is important that the precession constant $\alp$ is 
proportional to $H$ and inversely proportional to \OmE\
(see~\ref{sec:kbet}). Thus, because $H \ \propto \ \OmE^2$,
\beqn
\alp \ \propto \ \OmE \ .
\eeqn

Tidal dissipation refers to the energy dissipation in the earth 
and ocean, which reduces Earth's rotation rate and increases the 
length of day and the Earth-Moon distance. 
The parameter relevant here for PT solutions
is the change in lunar mean motion $n_L$ (average angular frequency), 
which is presently decreasing at a rate:
\beqn
Q_0 = (dn_L/dt)_0 / (n_L)_0 = -4.6 \e{-18} \ \mbox{s\pmo}
\eeqn
\citep{quinn91}. 
Reconstructing \Td's history 
involves considerable uncertainties and
remains an active area of research, specifically the
development of model frameworks for the Earth-Moon's tidal 
evolution focusing on
the long-term history of ocean tidal dissipation 
\citep[e.g.,][]{webb82,hansen82,kagan94,green17,
motoyama20,daher21,tyler21,farhat22,zeebelantink24pa}.

The combined precession-tilt-orbital solution \rev{may be 
denoted as OS(\Ed,\Td), see \citet{zeebelourens22pa}.}
For example, \ZBETa(1,1) refers to a PT solution 
based on the orbital solution \ZBETa\ with modern values 
$\Ed = 1$ and $\Td = 1$ (see Figs.~\ref{fig:cpfft} 
and~\ref{fig:oblqET}). Relatively simple
parameterizations using \Ed\ and \Td\ \rev{as employed
in \citet{zeebelourens22pa}} are most 
useful over, say, the Cenozoic, where changes in \PSI, Earth-Moon 
distance, etc.\ are comparatively small \citep{quinn91,zeebelourens22pa,
zeebe22aj}. In order to compute full PT solutions in deep time,
say, on Gyr-time scale, a fundamentally different approach is 
required that considers the long-term evolution of the 
Earth-Moon system, as well as the large scale dynamical chaos 
in the solar system \rev{\citep[see Sections~\ref{sec:psilong},
\ref{sec:deep} and][]{zeebelantink24aj,zeebelantink24pa}.}

\subsection{Up-to-date Precession and Obliquity Solutions
\label{sec:u2dPT}}

\begin{figure}[t]
\vspace*{-02ex} \hspace*{-02ex}
\includegraphics[width=1.0\linewidth]{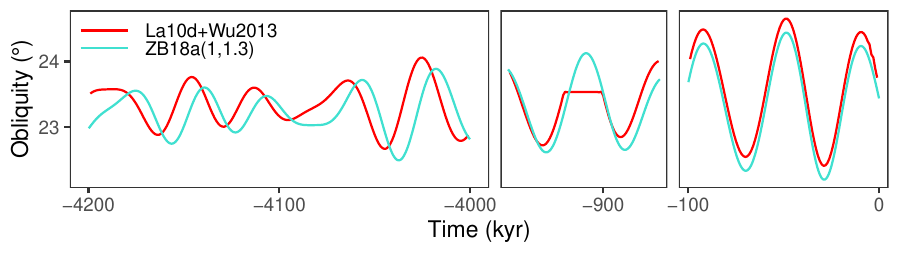}

\caption{\scs
Obliquity solution based on the procedure
by \citet{wu13ppp} for \LaXx\ currently available in the
\acl\ package \citep{liacycle19} (example: \LaXd+Wu2013,
red). For comparison, obliquity calculated
using \texttt{snvec} \citep{zeebelourens22epsl} based
on \ZBETa\ \citep{zeebelourens19} (\ZBETa(1,1.3),
turquoise, see text for details).
\label{fig:oblqWu}
}
\end{figure}

\citet{wu13ppp} suggested and employed a procedure to obtain 
obliquity and precession solutions using the orbital solutions 
\LaXx\ from \citet{laskar11} and the Fortran code 
from \citet{laskar93} (for an example following Wu~et~al.\
available in the \acl\ package \citep{liacycle19}, 
see Fig.~\ref{fig:oblqWu}).
\citet{wu13ppp} used the parameter values $\texttt{FGAM} = 
1.0$ and $\texttt{CMAR} = 1.3$.
Importantly, the \citet{wu13ppp} procedure/output has 
also been used in subsequent studies \citep[e.g.,][]{laurin15,
mameyers17,mameyers19}. However, the \LaXx\ output
is incompatible with the 1993-Fortran routine
because of different reference frames (see below). Here we 
illustrate the problematic obliquity solution by comparison 
with \citet{zeebelourens19}, from which we calculate obliquity 
using \texttt{snvec} \citep{zeebelourens22epsl} and the 
values for $E_{d} = 1$ and $T_{d} = 1.3$. 
The \acl\ output is 
offset already at $t = 0$ (see Fig.~\ref{fig:oblqWu}). 
Furthermore, at around $-900$~kyr, part of the cycle is clipped, 
which appears to be a numerical error. At around $-4$~Myr the 
\citet{wu13ppp} solution is offset from \citet{zeebelourens22epsl} 
by about half a cycle.

The problem with the approach of \citet{wu13ppp} is incompatible 
reference frames (see Section~\ref{sec:einc}).
The 1993-Fortran routines from \citet{laskar93} expect orbital 
elements given in the ecliptic reference frame. However, the \LaXx\
elements are given in the invariant reference
frame \citep{laskar11}. As explained above, the invariant frame is 
based on the invariable plane (perpendicular to the solar system's 
total angular momentum vector that passes through its barycentre)
and is hence different from the ecliptic frame. Wu~et~al.'s 
procedure is therefore not recommended. For the user interested
in PT solutions based on up-to-date orbital solutions, say, for
the Cenozoic, the freely available routines \snv\ and \snvR\
are recommended (\snvurl\ and \snvurlR). These routines can be 
used with up-to-date orbital solutions such as \ZBETa\ and \ZBXXx\
that supply orbital elements in a compatible reference frame.
For the application-oriented user, pre-computed and up-to-date 
PT solutions for different \Ed\ and \Td\ values are readily 
available for download (no programs, coding, etc.\ involved), 
see \npurlPT\ \citep{zeebelourens22pa}.
PT solutions here refer to full solutions (including 
individual PT cycles and temporal resolution of order 
$10^2$-$10^3$~y), rather than calculated frequencies and 
long-term averages.

\subsection{Milankovi{\'c} Forcing in Deep Time \label{sec:deep}}

\begin{figure}[t]
\vspace*{-40ex} \hspace*{-03ex}
\includegraphics[scale=0.8]{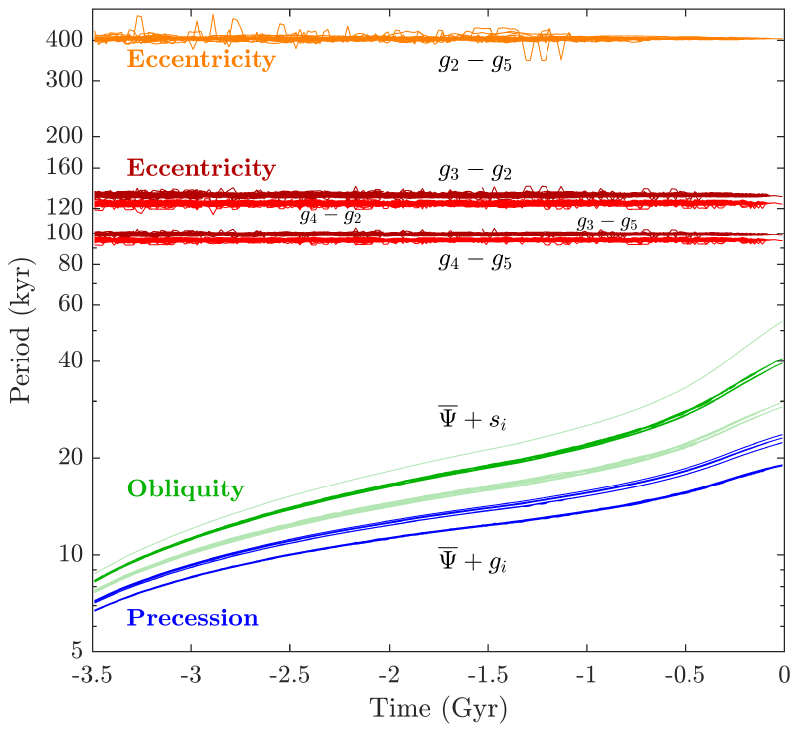}

\vspace*{-30ex}
\caption{\scs
Summary of eccentricity, tilt, and precession (ETP) periods from 
3.5-Gyr integrations \citep{zeebelantink24aj,zeebelantink24pa}.
Note the logarithmic ordinate.
\Psibar\ refers to the averaged luni-solar precession
over 20~Myr intervals.
\label{fig:AllT}
}
\end{figure}

This review has largely focused on the past few 100~Myr. 
Milankovi{\'c} forcing for older intervals (deep time) needs
to consider a host of additional factors, including tidal 
dissipation, the long-term evolution of the Earth-Moon system,
solar mass loss, and more. Unfortunately,
a comprehensive review of astronomical forcing
in deep time is beyond the scope of this paper.
Instead, we refer the reader to recent work specifically
dedicated to Milankovi{\'c} forcing in deep time 
\citep{zeebelantink24aj,zeebelantink24pa}, as well as 
related studies
\citep[e.g.,][]{macdonald64,goldreich66,mignard81,
berger89ast,ito93,touma94,ito95,waltham15,meyers18,spalding18,
farhat22,devleesch24,malinverno24}.
The evolution of the main periods of eccentricity, tilt,
and precession (ETP) on Gyr-time scale may be summarized in a 
nutshell as follows (see Fig.~\ref{fig:AllT}).
The average precession and obliquity periods ($\Psibar + g_i$
and $\Psibar + s_i$) increase with time due to the changing
luni-solar precession \Psibar\ (averaged over 
20~Myr intervals), accompanied by the slowing of Earth's 
rotation, lengthening of the day, and the receding Moon.
The short eccentricity periods (\sm{95}, 99, 124, 
and 131~kyr recent) appear relatively stable but 
actually exhibit substantial variations in frequency and 
amplitude and do not serve as metronomes 
\citep{zeebelantink24pa}. The spikes in the long
eccentricity cycle \gtfL\ (top, Fig.~\ref{fig:AllT})
reflect instabilities
in $g_2$, which compromises the 405-kyr cycle's 
reliability beyond several hundred million years in 
the past, as mentioned above \citep[][]{zeebelantink24aj}.
\rev{One major issue identified by \citet{zeebelantink24aj} 
is that the long eccentricity cycle may not be available 
as a primary tuning target in deep time.}

\rev{
Importantly, the direct use of astronomical solutions for astronomical 
tuning (e.g., for absolute ages and chronologies across the Cenozoic) 
represents only one realm of 
the use of astronomical solutions in astrochronology.
As pointed out in Section~\ref{sec:intro}, much of the ongoing 
astrochronological work on the geologic time scale uses
relative (floating) age models, spanning most of Earth's 
history, that is, the entire pre-Cenozoic. Pre-Cenozoic studies 
also rely on astronomical solutions in a range of capacities, even 
when explicit cycle-to-cycle tuning in the time domain is not possible 
due to uncertainties in amplitude, phase, and frequency of the forcing 
due to chaos and uncertainties in parameters such as tidal dissipation, 
dynamical ellipticity, etc. For more information
on pre-Cenozoic astronomical 
parameters, we refer the reader to recent theoretical work
\citep[e.g.,][]{spalding18,daher21,tyler21,farhat22,zeebelantink24aj,
zeebelantink24pa},
as well as observational work using the geologic record
\citep[e.g.,][]{zhang15orb,mameyers17,meyers18,kent18,
lantink19,olsen19,soerensen20,lantink22,lantink23,lantink24,
malinverno24,wumalinverno24}.
}

This section concludes our review of orbital solutions
and precession-tilt solutions. The remaining sections
deal with practical considerations, summary, and
outlook.

%
\section{Practical Considerations \label{sec:pract}}

In the following, we provide a few recommendations and 
list selected resources for practical consideration.
First, regarding orbital solutions, the user needs
to consider their valid time span, which is limited 
fundamentally by solar system chaos (see Table~\ref{tab:os}).
As pointed out above, the probability that a particular OS
represents the actual, unique history of the solar system is 
near zero significantly beyond the OS's validity range.
For example, finding a good match between, say an
eccentricity record at 100~Ma and La04 would be entirely 
coincidental and concluding that La04 is therefore a 
proper solution of the solar system orbits over the past 100 Myr 
would be misleading (see Section~\ref{sec:u2dOS}).
Instead, for the time being, we recommend the solutions 
\ZBETa\ and \ZBXXx\ across the time span from $-300$ to 0~Myr,
available at \npurlXXx\ and \myurl.
Note though that these solutions have been constrained by 
geologic data only up to an age of 66~Ma
(Table~\ref{tab:os}). Beyond $-300$~Myr, the \ZBXXtx\ 
solutions may serve as possible templates for deep time 
characteristics (see \npurlXXtx). Importantly, however, 
no single solution should be considered the ``true'' solution.

For up-to-date precession-tilt solutions, we recommend choosing
between two options at this time (see Section~\ref{sec:u2dPT}). 
One option is downloading 
pre-calculated PT solutions for different values of \Ed\ and \Td, 
available at \npurlPT\ and \myurl\ \citep[see][]
{zeebelourens22pa,zeebe22aj}. Another option is for the user to 
generate their own PT solutions based on up-to-date orbital 
solutions using our \snv\ code in C or R available at \snvurl\ 
and \snvurlR\ (for summary, see Table~\ref{tab:rsc}). We do not 
recommend the procedure proposed by \citet{wu13ppp} to generate 
PT solutions (see Section~\ref{sec:u2dPT}).
For the user interested in generating insolation curves
based on up-to-date orbital solutions, we mention
here a web-based user interface currently under construction
at \uiurl.

\subsection{Available Resources}

To assist the user in identifying available resources
for the analysis of their specific problem at hand, 
we list a few selected solutions, codes, and tools (including URLs,
see Table~\ref{tab:rsc}). The list is far 
from exhaustive and focuses on resources that are up-to-date
(including OS/PT solutions), 
open source, directly accessible, frequently used, and maintained. 
Assume the user's goal is, for example, to develop a 
cyclostratigraphic age model for a Miocene record based 
on astronomical tuning to Earth's precession and obliquity.
For the time series analysis of the data record, tools such as 
\ascr\ and \acl\ are available (Table~\ref{tab:rsc}). 
For the astronomical tuning to precession and obliquity,
PT solutions based on \ZBETa\ with different values for \Ed\ and 
\Td\ are available for download at \myurl\
\citep{zeebelourens22pa,zeebe22aj}. Given a task/time 
interval at hand and the information compiled in 
Tables~\ref{tab:os} and~\ref{tab:rsc}, most users
seeking guidance
should be able to identify resources for the analysis of their
problem using astronomical solutions and Milankovi{\'c} 
forcing as reviewed in the present paper. The experienced
user is of course free to employ the tools of their choice.
However, regarding astronomical solutions, we strongly recommend
considering the recency, limitations, and valid time span
of the solutions
(see Sections~\ref{sec:u2dOS} and~\ref{sec:u2dPT}).
Tools that yield inaccurate results should be abandoned
(see Section~\ref{sec:u2dPT}).

\renewcommand{\baselinestretch}{1.2}\selectfont
\begin{table}[h]
\caption{Selected resources (up-to-date OS/PT solutions, open source, 
directly accessible, frequently used, maintained).
\label{tab:rsc}
}
\vspace*{5mm}
\hspace*{-5mm}
\begin{tabular}{llll}
\hline
Resource Type             &  Name/ID    & Reference                & URL      \\
\hline
Orbital Solution          & \ZBXXtx         
                          & \scs\citet{zeebelantink24aj}           & $^{a,b}$ \\
Orbital Solution          & \ZBETa\ \ZBXXx\ 
                          & \scs\citet{zeebelourens19,zeebelourens22epsl}  & $^{c,a}$ \\
Orbital Solution          & \LaXx       & \scs\citet{laskar11}     & $^{d}$   \\
PT Solution               & \ZBETa(\Ed,\Td) 
                          & \scs\citet{zeebelourens22pa,zeebe22aj} & $^{e,a}$ \\
Insolation web-interface  & \myui       & \scs under construction  & $^{f}$   \\
                          
Orbit Integrator+Code (C) & \orb        & \scs\citet{zeebe23aj}    & $^{g,a}$ \\
PT Integrator+Code (C, R) & \snv\ \snvR & \scs\citet{zeebe22aj}    & $^{h,a}$ \\
Integrator+Code (C, Python) & \rbnd     & \scs\citet{reinliu12}    & $^{i}$   \\
Astrochronology Tool (R)  & \ascr       & \scs\citet{meyers14ascr} & $^{j}$   \\
Time Series Analysis Tool & \acl        & \scs\citet{liacycle19}   & $^{k}$   \\
\hline
\end{tabular}
\renewcommand{\baselinestretch}{\blss}\selectfont
\noindent {\scs \\[2ex]
$^a$ \myurl   \\
$^b$ \npurlXXtx \\
$^c$ \npurlETa\ $|$ \npurlXXx \\
$^d$ \LaXurl  \\
$^e$ \npurlPT \\
$^f$ \uiurl   \\
$^g$ \giturl  \\
$^h$ \snvurl\ $|$ \snvurlR \\
$^i$ \rbndurl \\
$^j$ \ascrurl \\
$^k$ \aclurl  \\
} 
\end{table}
\renewcommand{\baselinestretch}{\bls}\selectfont

%
\section{Summary \& Outlook \label{sec:summ}}

We have reviewed astronomical solutions and 
Milankovi{\'c} forcing as applied in the Earth sciences. Given
confusion and recent inaccurate results in the application
of astronomical solutions, our review appears appropriate and timely, 
and clarifies the astronomical basis, applicability, and 
limitations of astronomical solutions. Applying accurate and
up-to-date solutions and tools is critical to extend and improve
the astronomical time scale, which, over the past few decades, 
has transformed the dating of geologic archives.
Today, astronomical solutions and the astronomical time scale 
represent the backbone of astrochronology and cyclostratigraphy, widely 
used in the Earth sciences. In our review,
we have discussed two fundamental
types of astronomical solutions and their limitations: orbital solutions 
(OSs) and precession-tilt (PT) solutions. While OSs are a prerequisite
and essential for PT solutions, the reverse effects of rotational
dynamics on OSs are generally minor.

To understand the behavior of orbital solutions and astronomical
forcing terms, we have emphasized the analysis of the fundamental
(secular) frequencies, which may be thought of as the 
spectral building blocks of the dynamical system.
The secular frequency analysis also provides insight into 
amplitude and frequency modulation of orbital forcing signals 
and leads to the conclusion that 
the current evidence does not support the
notion of generally stable and prominent astronomical ``metronomes''
for universal use in astrochronology and cyclostratigraphy.
Large-scale dynamical chaos is an inherent characteristic of the 
solar system and fundamental to understanding the behavior of 
astronomical solutions and their limitations.
We have illustrated the unpredictability and unaccountability of 
chaotic systems in terms of basic statistics. One important
corollary is that it is generally not possible to ``tune'' or ``fit''
astronomical solutions to geological data.
\rev{The best that can probably be done is to create large ensembles 
of solutions via parameter variations and select/discard those
solutions that show good/poor agreement with the data.}
Our in-depth description of the various quantities associated 
with Earth's axial precession should aid in clarifying the 
terminology and notation appearing in the literature.
\rev{We have also discussed astronomical forcing and astrochronology 
in deep time, which represents another critical application 
of astronomical solutions beyond the direct use for astronomical 
tuning of, say, absolute Cenozoic ages and chronologies.
}
Finally, our summary of selected available resources
should assist the interested user in identifying resources for the 
analysis of their specific problem at hand, using the astronomical 
solutions and Milankovi{\'c} forcing as reviewed here.

Continuing improvements in computer technology, speed, and 
algorithms will accelerate numerical computations in the 
future and will allow integrating more complete and complex 
models of the solar system \cite[for recent endeavors, see, e.g.,]
[]{brownrein20,zeebe23aj,javaheri23,hernandez24}.
Importantly, however, the limits imposed by dynamical 
chaos on astronomical calculations and geological/astrochronological 
applications are fundamental and physical in nature, and cannot be 
overcome by numerical advances. 
In order to generate ensemble solutions with long (and specific)
divergence times, arbitrary precision code might be helpful.
One goal regarding user access to up-to-date
astronomical solutions in the future is offering more 
user-friendly applications, such as a web-based user interface
(for current efforts in progress/under construction, 
see, e.g., \uiurl). In terms of future observational endeavors, 
it is highly desirable to obtain high-quality
cyclostratigraphic sequences that
constrain the evolution of Earth's precession and obliquity
frequencies across the Cenozoic, as well as in deep time, 
the latter being critical to reconstructing the Earth-Moon 
system's long-term evolution. \rev{\citep[see, e.g.,][and 
references therein]{wumalinverno24}.} New
high-quality eccentricity and obliquity records older than
\sm{50}~Ma will be crucial in constraining the solar system's 
chaotic history, and in turn, in providing highly accurate 
geological dates and chronologies for cyclostratigraphy
and astrochronology.

%
\vspace*{2ex}
\noindent
{\bf Acknowledgments.}
We thank one anonymous reviewer, Dorian Abbot, Alberto Malinverno, 
and Steve Meyers for their thorough reviews, which improved the paper.
We are grateful to Alberto Malinverno and Steve Meyers 
for specific text suggestions that refined and extended the manuscript
content. We also thank David~M. Hernandez and Margriet Lantink for 
helpful comments on the manuscript draft.
This research was supported by Heising-Simons Foundation Grant 
\#2021-2800 and U.S. NSF grant OCE20-34660 to R.E.Z.

\vspace*{2ex}
software{
          \orb,         \url{github.com/rezeebe/orbitN},
          {\tt snvec},  \url{github.com/rezeebe/snvec},
          {\tt snvecR}, \url{github.com/japhir/snvecR}
          }



\begin{appendix}

{\small
\section{Precession \rev{Equations} \label{sec:kbet}}

The calculation of precession and obliquity using our \snv\ code 
is detailed in \citet{zeebe22aj} and \citet{zeebelantink24pa}. 
A few equations for the precession constant are included here 
for clarity (see Section~\ref{sec:pt}).
The change in the spin axis (unit vector \sv) is calculated from 
\citep[e.g.,][]{goldreich66,ward74,ward79,bills90,quinn91}:
\beqn
\dot{\sv} = \alp (\nv \cdot \nv) (\sv \x \nv) \ ,
\label{eqn:sdot}
\eeqn
where \alp\ is the precession constant \rev{(although 
cf.~Section~\ref{sec:psilong})}
and \nv\
the orbit normal (unit vector normal to the orbit plane). 
The obliquity (polar) angle, \obl, is given by:
\beqn
\cos \obl = \nv \cdot \sv \ .
\label{eqn:obl}
\eeqn
The precession (azimuthal) angle, $\prc$, measures the motion
of \sv\ in the orbit plane.
The precession constant \alp\ is calculated from 
\citep[e.g.,][]{quinn91}:
\beqn
\alp = K \ (\kap + \bet) \ ,
\label{eqn:alp}
\eeqn
where $\kap = (1 - e^2)^{-3/2}$ and $e$ is the orbital eccentricity.
$K$ and \bet\ relate to the torque due to the Sun and Moon,
respectively:
\beqn
K    & = & \q{3}{2} \q{C-A}{C} \q{1}{\OmE \ a^3} GM_S
\label{eqn:ksun} \\
\bet & = & g_L \ \q{a^3}{a_L^3} \q{m_L}{M_S} \ ,
\label{eqn:beta}
\eeqn
where $A$ and $C$ are the planet's equatorial and polar moments of 
inertia, $(C-A)/C = H$ is the dynamical ellipticity, \OmE\ is
Earth's angular speed, $a$ the semi-major axis of its orbit,
$a_L$ is the Earth-Moon distance parameter,
and $GM$ is the gravitational parameter of the Sun (see 
Table~\ref{tab:notval}). The index `$L$' refers to lunar properties,
where $g_L$ is a correction factor related to the lunar orbit 
\citep{kinoshita75,kinoshita77,quinn91} and $m_L/M_S$ is the lunar
to solar mass ratio. The parameter values used for Earth are given
in Table~\ref{tab:notval} and in \citet{zeebe22aj}. With $K$ and 
\bet, \alp\ and hence the luni-solar precession rate $\Psi$ can be 
calculated (see Eq.~(\ref{eqn:psi})).

}

\end{appendix}



\renewcommand{\baselinestretch}{0.0} 
\scs

\bibliographystyle{elsarticle-harv}

\end{document}